\documentclass[journal,draftcls,onecolumn,12pt,twoside]{IEEEtranTCOM}
\onecolumn

\usepackage{amsmath, amssymb,amsthm,cite}
\usepackage[dvips]{graphics}
\usepackage{graphicx}
\usepackage{psfrag}
\usepackage{bbm}
\usepackage{subfigure}
\usepackage{dblfloatfix}
\usepackage{comment}
\usepackage{mathtools}
\usepackage{algpseudocode}
\usepackage{algorithm}
\usepackage{hyperref}
\usepackage{color}
\usepackage{tcolorbox}
\usepackage{etoolbox}

\DeclareMathOperator*{\nn}{\nonumber}

\allowdisplaybreaks
\newcommand{\RNum}[1]{\uppercase\expandafter{\romannumeral #1\relax}}
\newtheorem{lemma}{Lemma}
\newtheorem{theorem}{Theorem}

\theoremstyle{definition}
\newtheorem{remark}{Remark}

\makeatletter
\def\blfootnote{\gdef\@thefnmark{}\@footnotetext}
\makeatother

\def\cX{{\mathcal X}}
\def\cY{{\mathcal Y}}
\def\cS{{\mathcal S}}
\def\cP{{\mathcal P}}
\def\cQ{{\mathcal Q}}
\def\cL{{\mathcal L}}
\def\cV{{\mathcal V}}
\def\vd{{\mathsf d}}
\def\todo{\textcolor{red}}
\def\archive{\textcolor{black}}

\def\patrick{\textcolor{black}}

\title{Graph-Based Encoders and their Performance for Finite-State Channels with Feedback}

\author{
\IEEEauthorblockN{Oron Sabag,}
\and
\IEEEauthorblockN{Bashar Huleihel, and}
\and
\IEEEauthorblockN{Haim H. Permuter}
}
\begin{document}
\maketitle
\vspace{-1.5cm}

\begin{abstract}
    The capacity of unifilar finite-state channels in the presence of feedback is investigated. We derive a new evaluation method to extract graph-based encoders with their achievable rates, and to compute upper bounds to examine their performance. The evaluation method is built upon a recent methodology to derive simple bounds on the capacity using auxiliary directed graphs. While it is not clear whether the upper bound is convex, we manage to formulate it as a convex optimization problem using transformation of the argument with proper constraints. The lower bound is formulated as a non-convex optimization problem, yet, any feasible point to the optimization problem induces a graph-based encoders. In all examples, the numerical results show near-tight upper and lower bounds that can be easily converted to analytic results. For the non-symmetric Trapdoor channel and binary fading channels (BFCs), new capacity results are eastablished by computing the corresponding bounds. For all other instances, including the Ising channel, the near-tightness of the achievable rates is shown via a comparison with corresponding upper bounds. Finally, we show that any graph-based encoder implies a simple coding scheme that is based on the posterior matching principle and achieves the lower bound.
\end{abstract}

\section{Introduction}\label{sec:intro}
\blfootnote{Part of this work was presented at the 2018 International Symposium on Information Theory (ISIT) \cite{OronBasharISIT}.}

\begin{figure}[t]
\centering
    \psfrag{E}[][][.95]{Encoder}
    \psfrag{D}[][][.95]{Decoder}
    \psfrag{C}[b][][.9]{$P_{Y|X,S}(y_t|x_t,s_{t-1})$}
    \psfrag{F}[t][][.9]{$s_t\mspace{-5mu}=\mspace{-5mu}f(y_t,x_t,s_{t-1})$}
    \psfrag{V}[][][.78]{Unit-Delay}
    \psfrag{M}[][][1]{$m$}
    \psfrag{Y}[][][1]{$y_t$}
    \psfrag{O}[][][1]{$\hat{m}$}
    \psfrag{Z}[][][1]{$y_{t-1}$}
    \psfrag{X}[][][1]{$x_t$}
    \includegraphics[scale = 0.55]{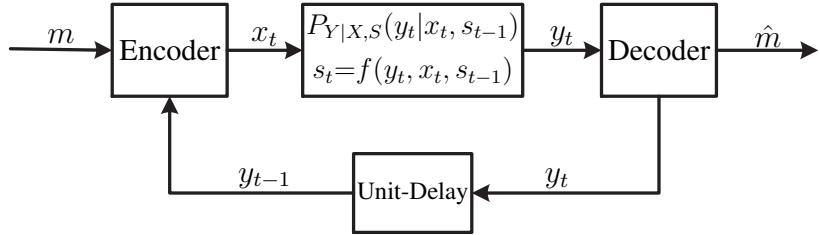}
    \caption{Unifilar FSC with feedback. The new channel state, $s_t$, is a function of $(y_t,x_t,s_{t-1})$.}
    \label{fig:FSC}
\end{figure}

Finite-state channels (FSCs) are commonly used to model scenarios in which the channel or the system have memory. Instances of this model can be found in wireless communication \cite{FSCTransWirelessComm,FSCTransWirelessComm1}, molecular communication \cite{MolecFSCTransComm,MolecularSurvey}, chemical interactions and magnetic recordings \cite{FSCMagnetic}. Despite their importance in theory and  practice, their capacity expression is still given by a non-computable  expression \cite{Gallager68,Loeliger_memory,GBAA,PfisterISI}. In this paper, we investigate computational methods for finding the capacity of unifilar FSCs with feedback (Fig. \ref{fig:FSC}).

A useful approach for computing the feedback capacity is via dynamic programming (DP) methods \cite{PermuterCuffVanRoyWeissman07_Chemical,Yang05,TatikondaMitter_IT09}. When the DP problem can be solved analytically, simple capacity expressions and optimal coding schemes can be determined \cite{Chen05,PermuterCuffVanRoyWeissman08,Ising_channel,Sabag_BEC,trapdoor_generalized,Ising_artyom_IT,PeledSabagBEC,Sabag_BIBO_IT}. However, in most cases analytical solutions are infeasible. Thus, no insights on communication aspects, such as coding schemes, or analytic expressions can be achieved, except to the resultant numerical lower bounds. In this paper, we propose an alternative method to compute lower and upper bounds on the capacity. The main advantages of the new evaluation method is that the numerical results can be converted into analytic expressions, and that each resultant lower bound implies a simple coding scheme.





\begin{figure}[t]
\centering
    \psfrag{Q}[][][0.95]{$Y=1$}
    \psfrag{E}[][][0.95]{$Y=0$}
    \psfrag{F}[][][0.95]{$Y=?$}
    \psfrag{O}[][][0.95]{$Y=0/?/1$}
    \psfrag{L}[][][1]{$Q=2$}
    \psfrag{H}[][][1]{$Q=1$}
    \includegraphics[scale = 0.5]{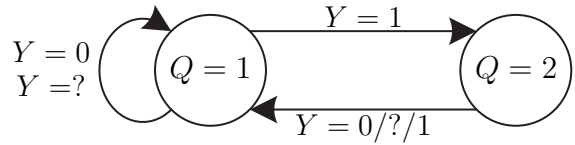}
    \caption{A $Q$-graph with $|\cQ|=2$ and $\cY=\{0,1,?\}$. The $Q$-graph is used to map channel output sequences onto unique node by walking along the labelled edges. For instance, if $Q=1$ and $Y=1$, then the graph will map this output to $Q=2$.}
    \label{fig:Q_intro}
\end{figure}

The upper and lower bounds are based on a new technique that simplifies the feedback capacity expression using auxiliary graphs \cite{Sabag_UB_IT}. The auxiliary graph, termed the \textit{$Q$-graph}, is used to map output sequences onto one of the auxiliary graph nodes (Fig. \ref{fig:Q_intro}). This sequential mapping can be exploited to derive single-letter lower and upper bounds on the capacity expression of the unifilar FSC \cite{Sabag_UB_IT}. Specifically, for any choice of a $Q$-graph, the upper bound is given by
\begin{align}\label{eq:UB_intro}
  C_\mathrm{fb}&\le \max_{P_{X|S,Q}} I(X,S;Y|Q),
\end{align}
where the joint distribution is $\pi_{S,Q}P_{X|S,Q}P_{Y|X,S}$, and $\pi_{S,Q}$ denotes a stationary distribution. For the lower bound, it was shown that any choice of a $Q$-graph yields
\begin{align}\label{eq:LB_intro}
  C_\mathrm{fb}&\ge I(X,S;Y|Q),
\end{align}
for all input distributions, $P_{X|S,Q}$, that are \textit{BCJR-invariant}, a property that will be defined later.

The upper bound plays an important role in capacity characterization as it is tight for all cases where the capacity is known, including the trapdoor, Ising and input-constrained channels. Furthermore, for all these cases, the upper bound is tight with auxiliary graphs that have small cardinality \cite{Chen05,PermuterCuffVanRoyWeissman08,Ising_channel,Sabag_BEC,trapdoor_generalized,Ising_artyom_IT,PeledSabagBEC,Sabag_BIBO_IT}. Therefore, if one can show a cardinality bound on the graph's size, it will suffice to conclude a single-letter capacity expression. The current paper was motivated by the question of whether there exists a cardinality bound or not. Unfortunately, we have no decisive answer to this question, but we developed very useful numerical tools that led to new capacity results, analytic bounds and simple coding schemes. 






First, we show that the upper bound in \eqref{eq:UB_intro} can be formulated as a standard convex optimization problem. The  convexity is not trivial since it depends on a stationary distribution that is controlled by the input distribution. As will be shown, the formulation gives efficient algorithms that converge to the global maximum that is required for the computation of \eqref{eq:UB_intro}. Second, given a conjectured solution, the upper bound can be proven analytically using the KKT conditions. The upper bound optimization problem is useful for evaluating the performance of the graph-based encoders that result from the lower bound optimization problem.

For the lower bound, we provide an optimization problem that maximizes the lower bound in \eqref{eq:LB_intro} over all BCJR-invariant input distributions. In this case, the optimization problem is not convex. Nonetheless, any feasible point (a BCJR-invariant input) induces a lower bound on the feedback capacity. The main advantage is that we can extract a $Q$-graph and an input distribution, termed here as a \emph{graph-based encoder} (in DP literature, a finite-state controller) and their achievable rates. Graph-based encoders also benefit from a simple coding scheme. We will present a posterior matching (PM) scheme that achieves the lower bound for any graph-based encoder. The scheme is inspired by the PM principle for memoryless channels \cite{shayevitz_posterior_mathcing} that was extended to systems with memory \cite{Sabag_BIBO_IT}. Thus, any graph-based encoder implies a simple coding scheme that achieves $I(X,S;Y|Q)$ even if the lower bound does not attain the capacity.

The optimization problems are formulated with respect to a fixed graph and evaluated with a generic enumeration method for directed graphs that we developed. An alternative method to constructing Markov graphs is also presented. These two construction methods are used to evaluate the bounds on well-known channels: the Ising, trapdoor and BFCs. The numerical results give promising results in all studies channels.

For all channels, graph-based encoders and their simple achievable rates are presented. The performance of the graph-based encoders, when compared to the numerical upper bounds, yield near-tight bounds. We also derive analytic upper bounds that lead to new capacity results. For example, for the BFC, we prove that the capacity is achieved with a graph-based encoder that has only a single node only. For the well-studied trapdoor channel, we derive a new capacity result by providing a simple graph-based encoder with only three nodes, and a corresponding upper bound.

The remainder of the paper is organized as follows: Section \ref{sec:prelimi} presents notation, the setting and background on the $Q$-graph bounds. Section \ref{sec:optimization} contains the optimization problems and the coding scheme. Section \ref{sec:analytic} contains examples, including their numerical evaluation and their analytic expressions. Lastly, Section \ref{sec:conclusion} contains some concluding remarks. Technical proofs are given in the appendices to preserve the flow of the presentation.

\section{Notation and Preliminaries}\label{sec:prelimi}
This section presents notation, the setting and the relevant background on the $Q$-graph \cite{Sabag_UB_IT}.
\subsection{Notation}
Random variables, realizations and sets are denoted by upper-case (e.g., $X$), lower-case (e.g., $x$) and calligraphic letters (e.g., $\mathcal X$), respectively. We use the notation $X^n$ to denote the $n$-tuple $(X_1,X_2,\dots,X_n)$ and $x^n$ to denote a realization of such a vector of random variables. For a real number $\alpha\in[0,1]$, we define $\bar{\alpha}=1-\alpha$. The binary entropy function is denoted by $H_2(\alpha) = -\alpha\log_2(\alpha)-\bar{\alpha}\log_2(\bar{\alpha})$. The cumulative distribution function of $X$ is denoted by $F_X$, and its inverse is denoted by $F_X^{-1}[w]=\min \{x:F_X(x)>w\}$. The probability vector of $X$ is denoted by $P_X$, the conditional probability of $X$ given $Y$ is denoted by $P_{X|Y}$, and the joint distribution of $X$ and $Y$ is denoted by $P_{X,Y}$. The probability $\Pr[X=x]$ is denoted by $P_X(x)$, and when the random variable is clear from the context, we write it in shorthand as $P(x)$. For a vector $\vd$, $\vd \preceq 0$ represents an element-wise inequality for each coordinate in $\vd$.

\subsection{FSC with feedback}
A FSC is defined by a conditional probability $P_{S^+,Y|X,S}$, where $X$ is the channel input, $Y$ is the channel output, $S$ is the channel state during transmission, and $S^+$ is the new channel state. The encoder chooses $x_t$, the channel input, based on the message $m$ and the output tuple $y^{t-1}$. At each time $t$, the channel has the property $P(s_t,y_t|x^t,s^{t-1},y^{t-1},m)=P_{S^+,Y|X,S}(s_t,y_t|x_t,s_{t-1})$. If the channel state, $S^+$, is a deterministic function $f(X,Y,S)$, then the FSC is called \textit{unifilar}. A unifilar FSC is \textit{strongly connected} if for all $s,s'\in\cS$, there exist $T$ and $\{P_{X_t|S_{t-1}}\}_{t=1}^T$ such that $\sum_{t=1}^T P_{S_t|S_0}(s|s') > 0$. It is also assumed that the initial state, $s_0$, is available to both the encoder and the decoder.

The capacity of the unifilar FSC is given by the following:
\begin{theorem}\label{theorem:capacity_unifilar}[Theorem $3$, \cite{PermuterCuffVanRoyWeissman08}]
The feedback capacity of a strongly connected unifilar FSC, where $s_0$ is available to both to the encoder and the decoder, can be expressed by
\begin{align}\label{eq:capacity_multiletter}
C_\mathrm{fb} &= \lim_{N\rightarrow \infty} \max_{\left\{P_{X_t|S_{t-1},Y^{t-1}}\right\}_{t=1}^N} \frac{1}{N} \sum_{i=1}^N   I(X_i,S_{i-1};Y_i|Y^{i-1}).
\end{align}
\end{theorem}

The capacity expression in Theorem \ref{theorem:capacity_unifilar} cannot be computed directly. It can be shown that the capacity can be formulated and evaluated as an infinite-horizon average reward MDP \cite{TatikondaMitter_IT09,PermuterCuffVanRoyWeissman08}. However, analytic solutions for the capacity are challenging due to the continuous alphabets of states and actions.
\subsection{The $Q$-graph bounds}\label{subsec:bounds}
The $Q$-graph bounds are an alternative for computing the capacity when the MDP cannot be solved. Their main idea is to simplify \eqref{eq:capacity_multiletter} by embedding an auxiliary graph into the capacity expression. We now formalize the $Q$-graph bounds that will be used in the optimization problems.

For an output alphabet $\mathcal{Y}$, the \textit{$Q$-graph} is a directed, connected and labeled graph. Each of its nodes should have $|\mathcal{Y}|$ outgoing edges with distinct labels (see an example in Fig. \ref{fig:Q_intro}).

The $Q$-graph definition implies that, given an initial node, $q_0$, and an output sequence, $y^t$, a unique node is determined by walking along the labelled edges according to $y^t$. The induced mapping can be represented by $\Phi_{t}:{\cY}^{t}\to \cQ$, or with a time-invariant function $g:\cQ\times\cY\to\cQ$, where a new graph node is computed from the previous node and the channel output.

Next, the $Q$-graph is embedded into the original FSC. A new directed graph, the \textit{$(S,Q)$-graph}, combines the $Q$-graph and the channel state evolution, and is constructed as follows:
\begin{enumerate}
  \item Each node in the $Q$-graph is split into $|\cS|$ nodes that are represented by pairs $(s,q)\in\cS\times\cQ$.
  \item An edge $(s,q)\rightarrow(s^+,q^+)$  with a label $(x,y)$ exists if and only if there exists a pair $(x,y)$ such that $s^+=f(s,x,y)$, $q^+=g(q,y)$, and $P(y|x,s)>0$.
\end{enumerate}
For a fixed $Q$-graph and distribution $P_{X|S,Q}$, the transition probabilities on the $(S,Q)$-graph are:
$$
P(s^+,q^+|s,q) = \sum_{x,y} P(x|s,q)P(y|x,s) \mathbbm{1}\{s^+ = f(s,x,y)\}\mathbbm{1}\{q^+ = g(q,y)\}.
$$
The notation $\mathcal{P}_{\pi}$ stands for the set of input distributions $P_{X|S,Q}$ that induce a unique stationary distribution on $(S,Q)$, that is, their corresponding $(S,Q)$-graph is irreducible and aperiodic.

Having defined the $(S,Q)$-coupled graph, the upper bound on the capacity can be presented:
\begin{theorem}\cite[Theorem $2$]{Sabag_UB_IT}\label{theorem:UB}
The feedback capacity of a strongly connected unifilar FSC, where the initial state is available both to the encoder and the decoder, is bounded by
\begin{align}\label{eq:Theorem_Upper}
C_{\text{fb}}\leq \sup_{P_{X|S,Q}\in\mathcal{P}_{\pi}}I(X,S;Y|Q),
\end{align}
for all $Q$-graphs for which the $(S,Q)$-coupled graph has a single and aperiodic closed communicating class. The joint distribution is $P_{Y,X,S,Q}=P_{Y|X,S}P_{X|S,Q}\pi_{S,Q}$, where $\pi_{S,Q}$ is the stationary distribution of the $(S,Q)$-coupled graph.
\end{theorem}

To present the lower bound, it is convenient to present the joint distribution as:
\begin{align}\label{eq:preliminari_joint}
P_{S,Q,X,Y,S^+,Q^+} = \pi_{S,Q}P_{X|S,Q}P_{S^+,Y|X,S}\mathbbm{1}\{Q^+ = g(Q,Y)\}.
\end{align}
The pairs $(S,Q)$ and $(S^+,Q^+)$ correspond to before and after a single transmission, respectively.

We define a property that is called \emph{BCJR-invariant input}. An input distribution $P_{X|S,Q}$ is said to be an \textit{aperiodic input} if its $(S,Q)$-graph is aperiodic. An aperiodic input distribution is BCJR-invariant if it implies the Markov chain:
$$
S^+ - Q^+ - (Q,Y),
$$
where the joint distribution is \eqref{eq:preliminari_joint}. A simple verification of the Markov chain is:
\begin{align}\label{eq:BCJR}
\pi_{S|Q}(s^+|q^+)
     &= \frac{\sum_{x,s} \mathbbm{1}_{\{s^+=f(y,x,s)\}}P_{Y|X,S}(y|x,s)P_{X|S,Q}(x|s,q)\pi_{S|Q}(s|q)}{\sum_{x',s'} P_{Y|X,S}(y|x',s')P_{X|S,Q}(x'|s',q)\pi_{S|Q}(s|q)},
\end{align}
which needs to hold for all $(s^+,q,y)$ and $q^+=g(q,y)$.

A \emph{graph-based encoder} is constituted of a $Q$-graph and a BCJR-invariant input distribution. The following theorem provides a lower bound on feedback capacity.
\begin{theorem}\cite[Theorem $3$]{Sabag_UB_IT}\label{theorem:lower}
The feedback capacity of unifilar FSCs is bounded by
\begin{align}\label{eq:Theorem_Lower}
C_{\text{fb}}&\geq I(X,S;Y|Q),
\end{align}
for all aperiodic inputs $P_{X|S,Q}\in\mathcal{P}_\pi$ that are BCJR-invariant.
\end{theorem}

\section{The optimization problems and coding scheme}\label{sec:optimization}
\patrick{The bounds on the feedback capacity (Theorems \ref{theorem:UB} and \ref{theorem:lower}) can be represented as follows:
\begin{align}\label{eq:bounds_intro}
  \max_{P_{X|S,Q}\in\cP_{BCJR}} I(X,S;Y|Q) &\le C_{\text{fb}} \le \max_{P_{X|S,Q}} I(X,S;Y|Q).
\end{align}
This section contains the formulation of two optimization problems, each corresponding to a bound in \eqref{eq:bounds_intro}, and the coding scheme. Note that the optimization problems only differ in their maximization domains. We will first provide a formulation of the upper bound as a convex optimization problem. Then, we introduce additional constraints that restrict the maximization domain to be on input distributions that are BCJR-invariant. The upper bound formulation and the extra constraints constitute the optimization problem of the lower bound in \eqref{eq:bounds_intro}.}


\subsection{The upper bound}
The optimization variables are chosen as $\vd\triangleq P_{S,Q,X,Y,S^+,Q^+}$, that is, a joint distribution on $\cS\times \cQ\times \cX\times \cY\times \cS\times \cQ$. The random variables $S$ and $S^+$ (correspondingly, $Q$ and $Q^+$) should be interpreted as the channel state (correspondingly, the $Q$-state) before and after one transmission. Thus, the optimization variables need to satisfy their original relation:
\begin{align}\label{eq:constraint_original}
  P_{S,Q,X,Y,S^+,Q^+}(s,q,x,y,s^+,q^+) &= P_{S,Q,X,Y}(s,q,x,y)\mathbbm{1}\{s^+ = f(s,x,y)\}\mathbbm{1}\{q^+ = g(q,y)\}.
\end{align}
With some abuse of notation, $\vd$ refers to joint distribution $P_{S,Q,X,Y,S^+,Q^+}$ that satisfies \eqref{eq:constraint_original}.

In the following, three sets of constraints for the optimization problem are defined:
\subsubsection{Stationary distribution} The random variables $(S^+,Q^+)$  are introduced to manipulate the joint distribution such that it has a stationary distribution on the $(S,Q)$-coupled graph. This is done by verifying that the marginal distributions satisfy $P_{S,Q}(s,q)=P_{S^+,Q^+}(s,q)$ for all $(s,q)$. 

Formally, for each $(s,q)$, the constraint function is given by
\begin{align}\label{eq:constraints_stationary}
f_i(\vd) &\triangleq \mspace{-10mu}\sum_{\{x,y,s^+,q^+\}}\mspace{-12mu}P_{S,Q,X,Y,S^+,Q^+}(s,q,x,y,s^+,q^+) - \mspace{-18mu} \sum_{\{s^-,q^-,x,y\}}\mspace{-12mu}P_{S,Q,X,Y,S^+,Q^+}(s^-,q^-,x,y,s,q),
\end{align}
where $i = 1,\dots,|\cS\times\cQ|$ is an index that corresponds to graph edges.
\subsubsection{Channel law} \patrick{The following set of constraint functions ensures that the distribution satisfies the Markov chain, $Y-(X,S)-Q$, and that the channel law is preserved. That is, $P_{Y|X,S}(y|x,s) = \frac{ P_{S,Q,X,Y}(s,q,x,y)}{\sum_{y'}P_{S,Q,X,Y}(s,q,x,y')}$ for all $(s,q,x,y)$. The corresponding constraint functions are given by}
\begin{align}\label{eq:constraints_channel}
f_i(\vd) &\triangleq P_{S,Q,X,Y}(s,q,x,y) - P_{Y|X,S}(y|x,s) \cdot \sum_{y'}P_{S,Q,X,Y}(s,q,x,y'),
\end{align}
for $i=|\cS\times\cQ|+1,\dots,|\cS\times\cQ| + |\cS\times\cQ\times\cX\times\cY|$. Note that $P_{Y|X,S}(y|x,s)$ is a constant that is given by the channel law.
\subsubsection{PMF} The last constraint function verifies that the optimization variables form a valid pmf,
\begin{align}\label{eq:constraint_PMF}
  f_K(\vd)&\triangleq \sum P_{S,Q,X,Y,S^+,Q^+}(s,q,x,y,s^+,q^+) - 1,
\end{align}
with $K= |\cS\times\cQ|(1 + |\cX\times\cY|)+1$.

In the following we define the optimization problem for the upper bound in Theorem \ref{theorem:UB}:
\begin{tcolorbox}[colframe=black,colback=white, sharp corners,colbacktitle=white,coltitle=black,boxrule=0.45pt]
\vspace{-2mm}
\underline{The optimization problem for the upper bound:}
\vspace{-0.3cm}
\begin{align}\label{opt:UB}
& \underset{\vd}{\text{minimize}} &f_0(\vd)& \triangleq -I(X,S;Y|Q)\nn\\
& \text{subject to} &f_i(\vd) & = \underline{0}, \; i = 1,\dots,K, \nn\\
& & -\vd & \preceq \underline{0},
\end{align}
\vspace{-1.2cm}

where $f_i(\vd)$ were defined in \eqref{eq:constraints_stationary}-\eqref{eq:constraint_PMF}.
\vspace{-2mm}
\end{tcolorbox}


The following theorem shows that \eqref{opt:UB} is a convex optimization problem.
\begin{theorem}[Convex optimization for UB]\label{theorem:UB_convex}
For a given $Q$-graph, the optimization problem in \eqref{opt:UB} is a convex optimization problem. That is, $f_i(\vd)$ are convex functions of $\vd$ for $i=0,\dots,K$.
\end{theorem}
The proof of Theorem \ref{theorem:UB_convex} appears in Appendix \ref{app:UB_convex}.
\patrick{Theorem \ref{theorem:UB_convex} is a computational result; the upper bound formulation as a convex problem makes it possible to use algorithms that converge to the global maximum, and are efficient in terms of running time.
For the implementation of Theorem \ref{theorem:UB_convex}, we used CVX \cite{cvx} with the Sedumi solver. Such simulations provide tolerances of $1e^-8$ for the objective and the constraints. This result complements the upper bound derivation in \cite{Sabag_UB_IT}, since it is now a \textit{computable} single-letter expression.
In Section \ref{sec:analytic}, we will illustrate the utility of the KKT conditions when simplifying the mutual information into analytic expressions.}
\begin{remark}
The natural choice for the optimization variables is the conditional distribution $P_{X|S,Q}$. This choice turned out to be challenging when attempting to show the objective convexity. The difficulty stems from the fact that the objective depends on the stationary distribution $\pi_{S,Q}$, which is an implicit function of $P_{X|S,Q}$. Specifically, the distribution $\pi_{S,Q}$ is given by the solution of $\underline{\pi}(T(P_{X|S,Q})-I) = \underline{0}$, where $T(P_{X|S,Q})$ is the transition matrix of the Markov chain $(S,Q)$ and $I$ is an identity matrix. Even for simple scenarios such as the entropy rate of a constrained Markov-chain \cite{Marcus98}, it is not clear whether the objective is convex, although it was observed numerically to behave as a convex function.
\end{remark}
\subsection{The lower bound}
In this section, we present the optimization problem of the lower bound. From a communication perspective, a $Q$-graph restricts the structure of cooperation between the encoder and the decoder. The idea behind the forthcoming optimization problem is to find the BCJR-invariant input distribution with the highest achievable rate when the structure of the cooperation (i.e., the $Q$-graph) is fixed\footnote{From an MDP perspective, the optimization problem looks for the best policy that is constrained to visit a finite number of states, subject to the $Q$-graph structure. Each node corresponds to an MDP state, and any path to this node should result in the same MDP state.}. The optimization problem for the lower bound is the upper bound in \eqref{opt:UB}, but with additional constraints. The constraints are imposed for the BCJR-invariant property:
\begin{equation}\label{eq:LB_BCJR}
  P_{S^+|Q,Y}(s^+|q,y) = P_{S^+|Q^+}(s^+|g(q,y)),
\end{equation}
for all $(s^+,q,y)$. Since $Q^+$ is a deterministic function of $(Q,Y)$, the constraint in \eqref{eq:LB_BCJR} can be viewed as the Markov chain $S^+-Q^+-(Q,Y)$.

Formally, for each $(s^+,q,y)$, the constraint function of the BCJR property in \eqref{eq:LB_BCJR} is:
\begin{align}\label{eq:constraints_BCJR}
	f_i(\vd) &\triangleq P_{S^+|Q,Y}(s^+|q,y) - P_{S^+|Q^+}(s^+|g(q,y))\nn\\
&= \frac{\sum_{s',x'} P_{S,Q,X,Y,S^+,Q^+}(s',q,x',y,s^+,g(q,y))}{\sum_{s',x',s''}P_{S,Q,X,Y,S^+,Q^+}(s',q,x',y,s'',g(q,y))} \nn\\& \quad- \frac{\sum_{s',q',x',y'} P_{S,Q,X,Y,S^+,Q^+}(s',q',x',y',s^+,g(q,y))}{\sum_{s',q',x',y',s''}P_{S,Q,X,Y,S^+,Q^+}(s',q',x',y',s'',g(q,y))},
\end{align}
where $i$ is an index that enumerates all triplets $(s^+,q,y)$ and takes values in $|\cS\times\cQ|(1 + |\cX\times\cY|)+2,\dots,K'$, where $K'=|\cS\times\cQ|(1+|\cY| + |\cX\times\cY|)+1$. One can already note that the constraints in \eqref{eq:constraints_BCJR} are not linear and, thus, the resulting optimization problem is not convex.

In the following we define the optimization problem for the lower bound:
\begin{tcolorbox}[colframe=black,colback=white, sharp corners,colbacktitle=white,coltitle=black,boxrule=0.45pt]
\vspace{-2mm}
\underline{The optimization problem for the lower bound:}
\vspace{-0.3cm}
\begin{align}\label{opt:LB}
& \underset{\vd}{\text{minimize}} &f_0(\vd)& \triangleq -I(X,S;Y|Q) \nn\\
& \text{subject to} &f_i(\vd) & = \underline{0}, \; i = 1,\dots,K, \nn\\
& & -\vd & \preceq \underline{0},
\end{align}
\vspace{-1.2cm}

where $f_i(\vd)$, $i=1,\dots,K'$ were defined in \eqref{eq:constraints_stationary}-\eqref{eq:constraint_PMF} and \eqref{eq:constraints_BCJR}.
\vspace{-2mm}
\end{tcolorbox}

For the implementation of \eqref{opt:LB}, we used a sequential quadratic programming (SQP) algorithm that is suitable for non-convex optimization problems. This method is implemented in MATLAB via a function called \textit{fmincon}. This function starts with an initial point for the solution, and then converges, possibly to the global maximum. Since the optimization problem is not convex, the termination point depends on the initial point. Therefore, we generate some random initial points and choose the solution that achieves the highest lower bound. Practically, we observed that for most $Q$-graphs, a few initial guesses are sufficient to converge to the global maximum.

The BCJR-invariant property was presented in \cite{Sabag_UB_IT} as a simple condition for the Markov chain $Y_{i}-Q_{i-1}-Y^{i-1}$ that in turn simplifies the capacity to the lower bound in \eqref{eq:Theorem_Lower}. It turns out that this property is also necessary when analyzing PM schemes \cite{Sabag_BIBO_IT} as will be shown next.

\subsection{Construction of graph-based coding schemes}
Each graph-based encoder, i.e., a feasible point to the optimization of the lower bound in \ref{opt:LB} benefits from the construction of an explicit matching coding scheme. We present the coding scheme construction with an informal statement of its achievable rate and discuss the missing (technical) details.

Throughout the scheme, a $Q$-graph and an input distribution $P_{X|S,Q}$ are fixed ahead of communication. The scheme is given by a simple procedure that is repeated $n$ times. In each procedure, both the encoder and the decoder will keep track of the posterior probability (PP):
\begin{align}\label{eq:PP}
  \lambda(m)&\triangleq P(m|y^{i-1}).
\end{align}
The PP corresponds to the decoder's belief regarding the message at time $i$, which is also available to the encoder from the feedback. To encode, we will use matching (described below) of the PP to an input distribution $P_{X|S=s,Q=q}$, where $(s,q)$ are determined as follows:
\begin{itemize}
    \item The graph node $Q=q$ is determined from the channel outputs $y^{i-1}$.
    \item The state $S=s$ is determined for each message separately. Recall that the channel state can be determined from $(m,y^{i-1})$. Therefore, for each $m$, one can compute $s(m)$ which corresponds to the channel state when assuming $M=m$.
\end{itemize}

We now present the transmission procedure with $m^\ast$ denoting the correct message.
\begin{tcolorbox}[colframe=black,colback=white, sharp corners,colbacktitle=white,coltitle=black,boxrule=0.45pt,]
\underline{Procedure in the coding scheme:}\\
 1. The encoder transmits
 \vspace{-0.3cm}
 \begin{align*}
 x(m^\ast) &= F_{X|S=s(m^*),Q=q}^{-1}[\Lambda(m^\ast)],
 \end{align*}
 \vspace{-0.4cm}
  \ \ \ \ where
  \begin{align*}
    \Lambda(m^\ast) &= \frac1{\pi_{S|Q}(s(m^*)|q)}\sum_{\{m< m^\ast: s(m)=s(m^*)\}} \lambda(m)
  \end{align*}
 3. The channel output $y$ is revealed\\
 4. The PP (of each message) is updated recursively as
\begin{align}\label{eq:PP_update}
\lambda^+(m) &= \frac{P_{Y|X,S}(y|F_{X|S=s(m),Q=q}^{-1}[\Lambda(m)],s(m))}{P_{Y|Q}(y|q)}\lambda(m)
\end{align}
 5. The graph node is updated:
\vspace{-1.3cm}

\begin{align*}
q^+ &= g(q,y)
\end{align*}
\vspace{-1.25cm}

 6. The state (of each message) is updated:
\begin{align*}
s^+(m) &= f(s(m),x(m),y)
\end{align*}
\vspace{-1.4cm}

\underline{Decoding (after $n$ times):}
\vspace{-1.3cm}

\begin{align*}
    \hat{m} = \arg\max \lambda(m).
\end{align*}
\end{tcolorbox}

The following theorem concludes the achievable rate of the scheme.
\begin{theorem}[Informal]\label{theorem:scheme_informal}
For any BCJR-invariant input, $P_{X|S,Q}$, the scheme achieves $I(X,S;Y|Q)$.
\end{theorem}
This theorem provides a coding scheme with low complexity that achieves $I(X,S;Y|Q)$. In the context of the current paper, its main contribution is that any feasible point to the optimization problem of the lower bound is accompanied by a coding scheme. Clearly, if the lower bound is tight, the scheme is capacity-achieving. It is interesting to note that the recursive computation of the message PP is preserved for channels with memory, but with a different update rule \eqref{eq:PP_update}.

The analysis of the coding scheme is omitted in this paper for the sake of brevity and due to the many technical details that are required to show Theorem \ref{theorem:scheme_informal} precisely\footnote{Specifically, there a need for dithering of the messages before the actual transmission of channel inputs. Also, we need to use a message splitting operation in order to maintain accurate behavior of the stationary distribution.}. In \cite{Sabag_BIBO_IT}, we presented a rigorous proof for the binary-input binary-output (BIBO) channel with input constraints where the state is $x^-$. By replacing $x^-$ with a general state, $s$, the analysis is identical Theorem \ref{theorem:scheme_informal} is proved.

\begin{table}[b]
\caption{Valid $Q$-graphs}
\vspace{-0.5cm}
\label{table:NO_graphs}
\centering
\begin{tabular}{|c|c|c|c|c|c|}
  \hline
  Graph size & 2 & 3  & 4    & 5     & 6 \\ \hline
  No. Graphs ($|\cY|=2$) & 5 & 50 & 4866 & 21126 & 655424 \\ \hline
  No. Graphs ($|\cY|=3$) & 27 & 2297 & 463548 & - & - \\
  \hline
\end{tabular}
\end{table}
\subsection{Choice of $Q$-graphs}\label{subsec:graphs}
So far, we presented two optimization problems for a fixed $Q$-graph. Here, we proceed with the development of a practical algorithm that computes the lower and upper bounds. The main challenge here is how to choose $Q$-graphs that will result in tight bounds. Below, we present several approaches to choose $Q$-graphs: all of them are applicable in both optimization problems.

\subsubsection{Graphs pool (GP)}
A \textit{valid} $Q$-graph is a directed graph that is aperiodic, i.e., connected and has period $1$. A brute-force method to find $Q$-graphs is to create a pool of all valid graphs. This is a combinatorial problem whose output increases sharply as the graph size increases. However, we developed an enumeration method for all graphs, so that we only need to save a list of indices and a simple function that returns the graph. A useful observation is that different labelling of the nodes on the $Q$-graph will result in the same graph structure, which gives an improvement of a factor $|\cQ|!$. In Table \ref{table:NO_graphs}, the number of valid graphs is listed for $|\cY|=2,3$.

\subsubsection{Markov graphs}

\begin{figure}[t]
\centering
    \psfrag{Q}[][][1]{$\;Y=1$}
    \psfrag{E}[][][1]{$\;\;Y=0$}
    \psfrag{L}[][][1]{$Q=\ '1'$}
    \psfrag{H}[][][1]{$Q=\ '0'$}
    \includegraphics[scale = 0.6]{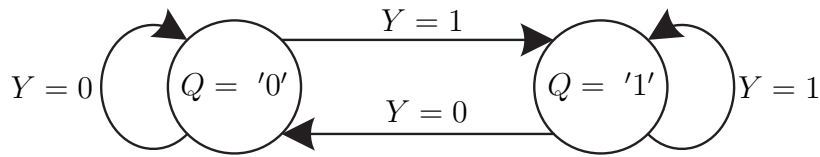}
    \caption{\patrick{A $Q$-graph where each node represents the last channel output ($k=1$) when $\mathcal{Y}=\{0,1\}$.}}
    \label{fig:1Markov}
\end{figure}

A valid choice of a $Q$-graph is a graph for which each node represents the last $k$ output symbols. For instance, see Fig. \ref{fig:1Markov}, where each node represents the last output from the alphabet $\cY=\{0,1\}$. For any choice of $k$ and an output alphabet $\cY$, the resultant graph has $|\cY|^k$ nodes and $|\cY|$ edges leaving each node.

Clearly, as $k$ increases, the performance of the bounds can be improved or unchanged. For several channels, it is known that the bounds will not approach the capacity for a finite $k$, since the optimal output distribution is a variable-order Markov process, which is a generalization of the Markov chain on the outputs that is suggested here.

\subsubsection{Discussion on \emph{continuous} graphs}
From a general perspective, we aim to solve the optimization problem
\begin{align}
    \min_{Q-\text{graphs}} \max_{P_{X|S,Q}} I(X,S;Y|Q).
\end{align}
This means that the upper bound is minimized over all applicable $Q$-graphs. This is a difficult problem since the minimization domain is discrete and, thus, should be searched through fully. A common technique in optimization is to relax a discrete domain into a continuous domain.

In our case, the relaxation is for the $Q$-graphs' domain in the minimization. Recall that, when $|\cQ|$ is fixed, a $Q$-graph function has the form $g: \cQ\times\cY\to\cQ$. To relax such a domain, note that such functions are exactly the boundaries of a conditional distribution $P_{Q^+|Q,Y}$. We will now show that the upper bound is also valid on the interior of $P_{Q^+|Q,Y}$.
\begin{lemma}[Upper bound with probabilistic $Q$-graph]
 For any $P_{Q^+|Q,Y}$, the upper bound in Theorem \ref{theorem:UB} holds.
\end{lemma}
The only difference from Theorem \ref{theorem:UB} is the transition law of the Markov chain on $(S,Q)$:
\begin{align}\label{eq:transition_continuous}
  P(s^+,q^+|s,q)&= \sum_{x,y}  \mathbbm{1}_{\{s^+=f(y,x,s)\}}P_{Q^+|Q,Y}(q^+|q,y)P_{Y|X,S}(y|x,s)P_{X|S,Q}(x|s,q).
\end{align}
\begin{proof}
From the functional representation lemma \cite{ElGamal}, $Q^+$ is a function of $(Q,Y,W)$, where $W$ is independent of $(Q,Y)$. We now define an auxiliary unifilar FSC with output $(Y,W)$, with $W$ independent of the input and the channel state. Clearly, the capacity of the new channel is the same as that of the original one, but now the $Q$-graph is labelled with $(Y,W)$, as needed.
\end{proof}

The objective of the corresponding optimization problem is now:
\begin{align}\label{eq:obj_minmax}
  &\min_{P_{Q^+|Q,Y}}\max_{P_{X|S,Q}} I(X,S;Y|Q).
\end{align}
However, it is not difficult to show that \eqref{eq:obj_minmax} can be formalized as a concave optimization problem when $P_{X|S,Q}$ is fixed. Thus, the optimal $Q$-graph lies on the boundaries of $P_{Q^+|Q,Y}$, that is, the optimal $Q$-graph is deterministic. This fact makes this relaxation attempt counterproductive.

\section{Examples and Analytic results}\label{sec:analytic}
In this section, we provide explicit graph-based encoders and prove their tightness when possible. In all other cases, we compare the achievable rates with numerical upper bounds.
For all the examples in this section, the variables take values from a binary alphabet, i.e., $\mathcal{S}=\mathcal{X}=\mathcal{Y}=\{0,1\}$.

\begin{figure}[t]
\centering
    \psfrag{A}[][][1]{$X_t$}
    \psfrag{B}[][][1]{$S_{t-1}$}
    \psfrag{C}[][][1]{$N_t\sim \mathsf{Ber}(p)$}
    \psfrag{C}[][][1]{$N_t$}
    \psfrag{D}[][][1]{$\;Y_t$}
    \includegraphics[scale = 0.9]{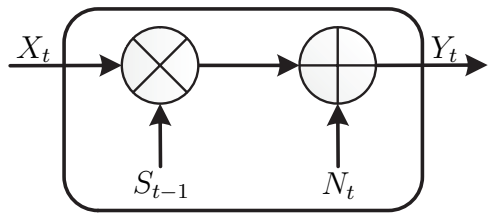}
    \caption{The BFCs. The channel input $x_t$ is multiplied by the channel state $s_{t-1}$, and then a XOR operation is performed with an i.i.d. sequence $N_t$ that is distributed according to $\text{Ber}(p)$.}
    \label{fig:Fading}
\end{figure}
\subsection{Binary fading channels} \label{sec: BFCs}
The binary fading channel (BFC), depicted in Fig. \ref{fig:Fading}, is a simplified version of the fading channel:
\begin{align} \label{eq: Fading_Channel}
Y_t = (S_{t-1} \cdot X_t)\oplus N_t,
\end{align}
where $\oplus$ is the XOR operation and $N_t$ is an i.i.d. sequence that is distributed according to $\text{Ber}(p)$ and is independent of the message. We study the following two scenarios of state evolution:
\begin{enumerate}
    \item Type I: Channel state evaluation is given by $S_t=S_{t-1}\oplus X_t\oplus Y_t$.
    \item Type II: Channel state evaluation is given by $S_t=S_{t-1}\oplus N_t$.
\end{enumerate}
Note that these are unifilar FSCs since $N_t$ is a function of $X_t$, $Y_t$ and $S_{t-1}$.
\subsubsection{The BFC of type I}

\begin{figure}[h]
\centering
        \psfrag{A}[][][1]{Channel parameter - $p$}
		\psfrag{B}[][][1]{Rate [bits/symbol]}
		\psfrag{C}[][][1]{BFC of type I - Upper and Lower bounds}
    \includegraphics[scale=0.38]{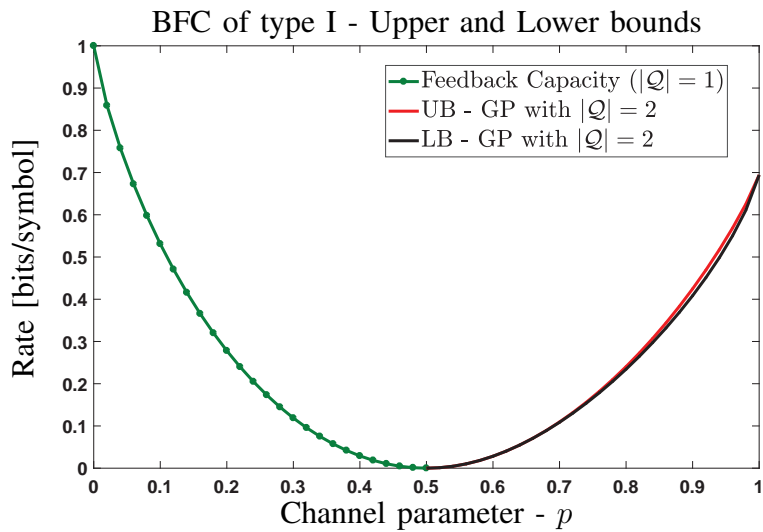} 
    \caption{Bounds on the capacity of the BFC of type $I$. Both the upper and lower bounds are obtained by using GP of size $|\cQ|=2$, while the feedback capacity is obtained by using a $Q$-graph with a single node.}
    \label{fig:fading1}
\end{figure}

The numerical evaluation is presented in Fig. \ref{fig:fading1}. First, for $p\le0.5$, the numerical evaluation showed that the upper and lower bounds are tight with a single node $Q$-graph. For $p\ge0.5$, upper and lower bounds that are based on the GP with $|\cQ|= 2$ are presented. The maximal difference between these bounds is $\sim0.02$, which means that the derived graph-based encoder achieves at least $96\%$ of the capacity. Finally, for the extreme point when $p=1$, the capacity is $\log_2(\phi)$ ($\sim0.694)$, where $\phi$ is the golden ratio. This follows by noting that if the channel is at state $s_{t-1}=0$, the encoder prefers to choose $x_t=0$ since it yields that the next channel state is $s_t=1$. 

The results are formalized in the following Theorem:
\begin{theorem}\label{theorem:Type1}
For the BFC of type I:
	\begin{enumerate}
		\item There exists a graph-based encoder with $|\cQ|=1$ that, for $p\le0.5$, achieves
		\begin{align}\label{eq:fading_encoder1}
			R_1^\mathrm{I}(p) = 1 - H_2(p).
		\end{align}
		\item The encoder in \eqref{eq:fading_encoder1} is optimal, that is, $C^\mathrm{FC}(p) = R_1^\mathrm{I}(p)$ for all $p\le0.5$.
		\item There exists a graph-based encoder with $|\cQ|=2$ that, for $p\in[0.5,1]$, achieves
\begin{align}\label{eq:fading_encoder2}
	R_2^\mathrm{I}(p) = \max_{a\in[0,1]} \; & \frac{1}{1+a\bar{p}+\bar{a}p}\bigg[H_2(a\bar{p}+\bar{a}p)  + (a\bar{p}+\bar{a}p)H_2\left(\frac{a\bar{p}}{a\bar{p}+\bar{a}p}\right)\bigg] - H_2(p).
\end{align}
	\end{enumerate}
\end{theorem}
\begin{figure}[b]
\centering
    \psfrag{C}[][][1]{}
    \psfrag{D}[][][1]{$Y=1$}
    \psfrag{E}[][][1]{}
    \psfrag{F}[][][1]{$Y=0$}
    \psfrag{G}[][][1]{$Y=0/1$}
    \psfrag{B}[][][1]{$Q=2$}
    \psfrag{A}[][][1]{$Q=1$}
    \includegraphics[scale = 0.5]{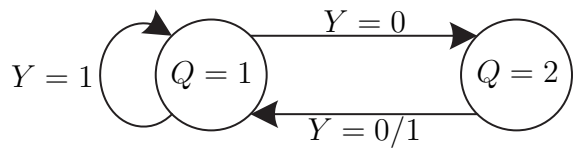}
    \caption{The $Q$-graph which is used to obtain the graph-based encoder for the BFC of type I.}
    \label{fig:FC_I_Q2}
\end{figure}
The proof of Theorem \ref{theorem:Type1} is given in Appendix \ref{app:type1}. For $p\le0.5$, the capacity result is surprising since an optimal single-node encoder implies that the optimal output distribution is i.i.d. To the best of our knowledge, this is the first instance of such a phenomena for channels with memory. For $p\ge 0.5$, the lower and upper bounds can be improved by using the GP with a larger graph size so that a graph-based encoder achieves at least $99\%$ of the capacity. However, we present this encoder in order to keep the achievable rate and the proof ideas simple. The $Q$-graph that is used for obtain the graph-based encoder of Theorem \ref{theorem:Type1} is presented in Fig. \ref{fig:FC_I_Q2}.

\subsubsection{The BFC of type II}

\begin{figure}[t]
\centering
        \psfrag{A}[][][1]{Channel parameter - $p$}
		\psfrag{B}[][][1]{Rate [bits/symbol]}
		\psfrag{C}[][][1]{BFC of type II - Upper and Lower bounds}
    \includegraphics[scale=0.38]{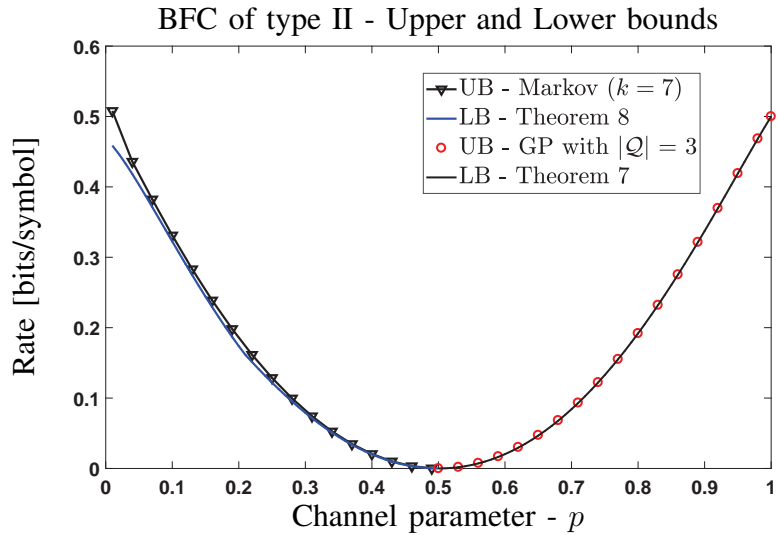}
    \caption{Bounds on the capacity of the BFC of type II. For $p\le0.5$, the lower bound is a maximum between two graph-based encoders of size $6$, and the upper bound is obtained using a Markov graph of size $|\cQ|=2^7$. For $p\ge0.5$, the lower bound is based on a graph-based encoder of size $3$ that is shown to be optimal for $p\ge p^*\sim0.75$.}
    \label{fig:fading2}
\end{figure}

The numerical evaluation for the BFC of type II is presented in Fig. \ref{fig:fading2}. For $p\ge0.5$, the lower bound is attained with a single graph of size $3$. The numerical upper bound shows that the lower bound is tight and we can prove it for $p\ge p^*$, where $p^*\sim0.75$.
\begin{theorem}\label{theorem:Type2_1}
For the BFC of type II,
\begin{enumerate}
\item There exists a graph-based encoder with $|\cQ|=3$ that, for $p\ge0.5$, achieves
		\begin{align*}
			R_1^\mathrm{II}(p) = \frac{1+H_2(2p\bar{p})-2H_2(p)}{2(2-p)}.
		\end{align*}
\item The feedback capacity is $$C_{\mathrm{II}}(p)= R_1^\mathrm{II}(p),$$ for $p\in[p^*,1]$, where $p^*$ is the unique solution for $(2 + p)\log_2\left(\frac{2p^2}{1 - 2 p\bar{p}}\right) +
 \log_2\left(\frac{2\bar{p}^2}{1 - 2 p\bar{p}}\right)=0$.
\end{enumerate}
\end{theorem}
The proof of Theorem \ref{theorem:Type2_1} appears in Appendix \ref{app:Type2_1}. The interesting part in the proof of Theorem \ref{theorem:Type2_1} is the upper bound. The maximizer of the upper bound can be intuitively guessed, but its proof is non trivial. To overcome this, we utilize the KKT conditions as sufficient and necessary conditions for the optimality of our initial guess.

For $p\le0.5$, we could not find capacity results. However, a near-tight lower bound is  attained using GP with $|\cQ|=6$, while the upper bound is evaluated with a Markov graph of size $|\cQ|=2^7$.

\archive{In Appendix \ref{app:Type2_1}, we present two graph-based encoders for the BFC of type II, for which the maximum between their achievable rates is the lower bound in Fig. \ref{fig:fading2}. The first graph-based encoder outperforms the other for $p\in[0,0.21]$ and vice versa}.



\subsection{The Ising channel}
\begin{figure}[b]
\centering
    \psfrag{A}[][][1]{Channel parameter - $p$}
		\psfrag{B}[][][1]{Rate [bits/symbol]}
		\psfrag{C}[][][1]{Ising channel - Upper and Lower bounds}	
    \includegraphics[scale=0.35]{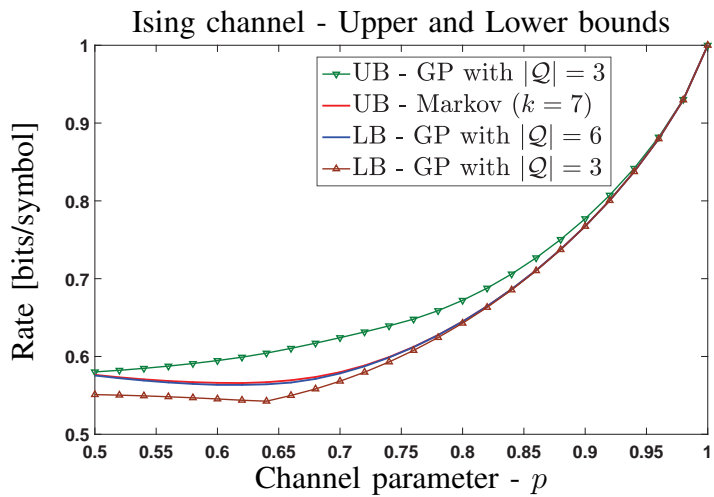}
    \caption{Upper and lower bounds on the capacity of the Ising channel. The red and blue solid lines represent upper and lower bounds, respectively; these almost coincide, such that the maximal difference is $\sim10^{-3}$.}
    \label{fig:ISING}
\end{figure}
The Ising channel was introduced by Berger and Bonomi \cite{Berger90IsingChannel} as a communication channel that resembles the physical Ising model. When the channel state is $s_{t-1}$ and the encoder transmits $x_t$, then, $y_t = s_{t-1}$ with probability $p$ or $y_t = x_t$ with probability $1-p$. The next channel state is $s_t=x_{t}$. The feedback capacity of this channel was derived for $p\le0.5$ in \cite{Ising_channel, Ising_artyom_IT}.


In Fig. \ref{fig:ISING}, we present the numerical evaluation when $p\ge0.5$. To emphasize the importance of the $Q$-graph construction, both methods are plotted. The upper and lower bounds, with triangle marks, are obtained using the GP with $|\cQ|=3$. The upper bound can be improved significantly with a Markov graph with $k=7$, that is, $|\cQ|=2^7$ (red solid line). The lower bound can also be improved by using the GP with $|\cQ|=6$ (blue solid line). The difference between the improved bounds is negligible and has an order of $\sim10-3$.

The improved lower bound (blue solid line) is attained as a maximum between two $Q$-graphs of size $6$, where each of them is superior for different channel parameters. Their difference is negligible, and we now formalize its achievable rate.

\begin{theorem}\label{theorem:Ising}
For the Ising channel, there exists a graph-based encoder with $|\cQ|=6$ that, for $p\ge0.5$, achieves
\begin{align}\label{eq:ising_FSE}
R_\mathrm{ISING}(p)  = \max_{(a,b,c)\in\mathcal{L}}\; &\frac{\bar{c}}{\bar{c}(1+a) + 1 - ab}\bigg[H_2(a) + a\cdot H_2(b)    + \frac{1 - ab}{\bar{c}}\cdot H_2(c) \nonumber
   \\& + \frac{(1 - ab)(k- 2p^3 + 2abp\bar{p}^2)}{k\bar{c}} \cdot H_2(p)\bigg],
\end{align}
where $k \triangleq 1-\bar{p}[(1+2p^2)(1-ab)+a\bar{c}\bar{p}+cp+3abp]$, and $\mathcal{L}$ denotes all $(a,b,c)$ that satisfy:
\begin{align}\label{eq:ising_L}
	0 &\leq c \leq 0.5 \leq a \leq b \leq 1 \nn\\
	0 &\leq p(1- ab) - \bar{a}b \nn\\
	0 &\leq a\bar{p}(\bar{b}- c)  - p\bar{c}  + abp + p^2(1 - ab) \nn\\
	0 &\leq ab\bar{p}(p^2+\bar{p}\bar{b}+\bar{c})+\bar{c}(p\bar{b}-a\bar{p})-p^2\bar{p}.
\end{align}
\end{theorem}

\archive{The proof of Theorem \ref{theorem:Ising} is presented in Appendix \ref{app:Ising}.}

\subsection{The trapdoor channel}
\begin{figure}[t]
\centering
    \psfrag{B}[][][1]{$1$}
    \psfrag{C}[][][1]{$0$}
    \psfrag{D}[][][1]{$1$}
    \psfrag{E}[][][1]{$0$}
    \psfrag{F}[][][1]{$1$}
    \psfrag{G}[][][1]{$1$}
    \psfrag{J}[][][1]{$x_{t+2}$}
    \psfrag{K}[][][1]{$x_{t+1}$}
    \psfrag{L}[][][1]{$x_t$}
    \psfrag{M}[][][1]{$s_{t-1}$}
    \psfrag{N}[][][1]{$y_{t-1}$}
    \psfrag{O}[][][1]{$y_{t-2}$}
    \psfrag{Q}[][][1]{Channel}
    \includegraphics[scale = 1]{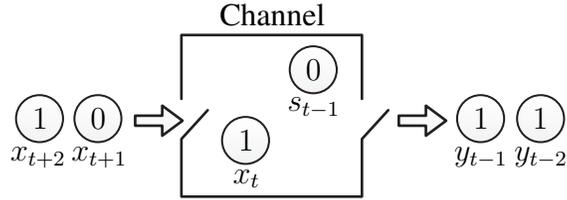}
    \caption{The trapdoor channel: at time $t$, the channel starts with a ball $s_{t-1}$ already in it, and a new ball $x_t$ is inserted into the channel. The channel output $y_t$ is $s_{t-1}$ with probability $p$ or  $x_t$ with probability $1-p$. The new state is the remaining ball.}
    \label{fig:Trapdoor_setting}
\end{figure}
The trapdoor channel was introduced by Blackwell as a \textit{``two-state simple channel"} \cite{Blackwell_trapdoor}. The channel is described in Fig. \ref{fig:Trapdoor_setting}. When the channel state is $s_{t-1}$ and the encoder transmits $x_t$, then, $y_t = s_{t-1}$ with probability $p$, and equals $y_t = x_t$ with probability $1-p$. The next channel state is computed as $s_t=s_{t-1}\oplus x_t\oplus y_t$, where $\oplus$ represents the XOR operation. The feedback capacity of this channel is still an open problem, except for $p\in[0.5,0.7035]$ \cite{PermuterCuffVanRoyWeissman08,trapdoor_generalized}.

\begin{figure}[t]
\centering
        \psfrag{A}[][][1]{Channel parameter - $p$}
		\psfrag{B}[][][1]{Rate [bits/symbol]}
		\psfrag{C}[][][1]{Trapdoor channel - Upper and Lower bounds}
    \includegraphics[scale=0.38]{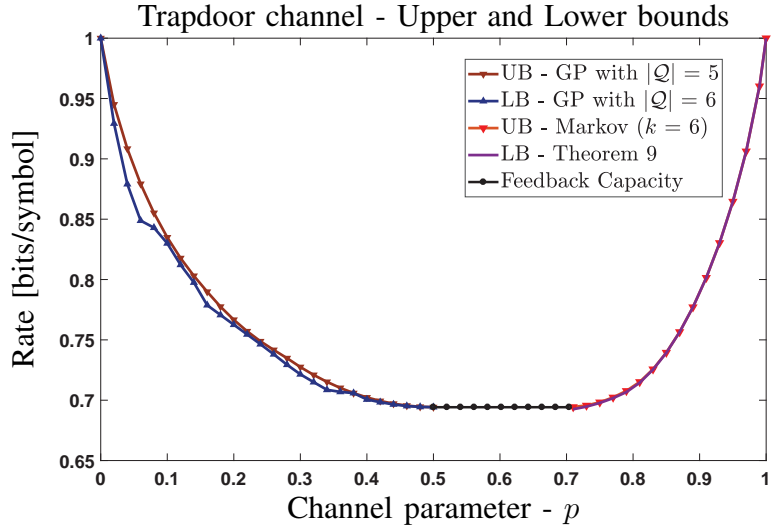}
    \caption{A comparison between upper and lower bounds on the capacity of the trapdoor channel.}
    \label{fig:Trapdoor}
\end{figure}

In Fig. \ref{fig:Trapdoor}, we present the numerical results for the trapdoor channel. For $p\le 0.5$, the upper bound is obtained using the GP  with $|\cQ|=5$, while for $p\ge 0.5$, we used a Markov graph with $k=6$. For $p\le 0.5$, the lower bound is attained as the maximum over eight particular $Q$-graphs of size $6$. For $p\in[0.5,0.7035]$ the feedback capacity is presented and for $p\ge 0.7035$ the graph-based encoder in Theorem \ref{theorem:TRAPDOOR}, denoted by $R_\mathrm{T_2}(p)$, is presented.

The following Theorem concerns the analytical formulation of the trapdoor channel results. We also prove a new capacity result by showing that the bounds are tight at a certain value of the channel parameter.

\newpage
\begin{theorem}\label{theorem:TRAPDOOR}
For the trapdoor channel,
\begin{enumerate}
  \item There exists a graph-based encoder with $|\cQ|=3$ that, for all $p\in[0,0.5]$, achieves:
  \begin{align}\label{eq:trapdoor_lower}
  R_\mathrm{T_1}(p) &= \frac{H_2(z) + z(1-H_2\left(\frac{1-z}{2-z}\right))}{1+z} ,
\end{align}
where $z=z(p)\triangleq \frac{1-2p}{1-p}$.
  \item There exists a graph-based encoder with $|\cQ|=4$ that, for all $p\in(0.5,1]$, achieves:
  \begin{align}\label{eq:trapdoor_lower2}
  R_\mathrm{T_2}(p) &= \max_{(a,b)\in[0,1]^2}\frac{1}{\bar{a}+\bar{b}}\left[\bar{b}H_2(a)+\bar{a}H_2(b) - \frac{\bar{a}(2p - 1)(p-\bar{a}\bar{p}-\bar{b}\bar{p})}{k}H_2(p)\right],
	\end{align}
where $k = 1 - \bar{p}(2a - 3ap + 5p + bp - 1)$,
and the maximum is over all $(a,b)$ that satisfy
\begin{align*}
	\frac{bp^2+a\bar{p}^2-\bar{p}p}{k} &\leq 0.
\end{align*}
  \item For all $p$, the feedback capacity of the trapdoor channel is upper bounded with:
\begin{align}\label{eq:Trapdoor_UB}
        C_\mathrm{T}(p)&\leq \max_a \frac{H_2(a) + a(1-H_2(p))}{1+a}.
\end{align}

  \item There exists a unique $p^*\in[0,0.5)$ such that the bounds in \eqref{eq:trapdoor_lower} and \eqref{eq:Trapdoor_UB} coincide. Therefore, at this point, the capacity is equal to $C_\mathrm{T}(p^*) = R_T(p^*)$.
\end{enumerate}
\end{theorem}

\archive{The proof of Theorem \ref{theorem:TRAPDOOR} is given in Appendix \ref{app:trapdoor}.}

\section{Conclusions and future work}\label{sec:conclusion}
We provided two optimization problems for computing the lower and upper bounds on the capacity of a unifilar FSC. The upper bound optimization problem is convex, while the lower bound optimization problem is non-convex. Nonetheless, we showed that the non-convex problem is still beneficial for the extraction of graph-based encoders that result in near-tight bounds. A direct relation between graph-based encoders and a PM scheme has also been established. Analytic results have been provided for several channels to demonstrate the efficiency of our computational tools.

The current work is part of the ongoing progress on the capacity of unifilar FSCs. Currently, it is unknown whether there exists a cardinality bound on the auxiliary random variable $Q$, that is, a bound on the graph's size. The notion of graph-based encoders can also be used in other communication scenarios with sequential nature, such as the two-way channel \cite{Sabag_TWC}.

\appendices
\section{Convexity of the upper bound --- Proof of Theorem \ref{theorem:UB_convex}}\label{app:UB_convex}
The optimization problem is convex if the objective is a convex function of $\vd$ and the constraints are linear functions of $\vd$.
 The inequality constraint $-\vd\preceq0$ and the equality constraints in \eqref{eq:constraints_stationary} - \eqref{eq:constraint_PMF}, are linear functions of $\vd$. Therefore, all that remains is to show the objective convexity.

The objective equals $ -H(Y|Q) + H(Y|X,S,Q)$. First, $H(Y|X=x,S=s,Q=q)$ is constant (given by the channel law), which implies that $H(Y|X,S,Q)$ is a linear function of $\vd$.

The second term is:
\begin{align*}
  -H(Y|Q) &= \sum_{q,y} \left(\sum_{x',s'}P(y,x',s',q)\right)\cdot \log \frac{\sum_{\hat{x},\hat{s}}P(y,\hat{x},\hat{s},q)}{\sum_{y,x,s}P(y,x,s,q)}.
\end{align*}

For each $(q,y)$, define
\begin{align}\label{eq:logsum_var}
    a_1 &= \lambda (\sum_{x',s'}P_1(y,x',s',q)), \quad a_2 = (1-\lambda) (\sum_{x',s'}P_2(y,x',s',q)),\nn\\
    b_1 &= \lambda (\sum_{y,x,s}P_1(y,x,s,q)), \quad b_2 = (1-\lambda) (\sum_{y,x,s}P_2(y,x,s,q)).
\end{align}

The log-sum inequality gives that:
$$a_1\log\frac{a_1}{b_1} + a_2\log\frac{a_2}{b_2} \ge (a_1+a_2) \log \frac{a_1 + a_2}{b_1 + b_2},$$
and substituting \eqref{eq:logsum_var} gives that for each $(q,y)$
\begin{align*}
& \lambda P_1(y,q)\log\left(\frac{P_1(y,q)}{P_1(q)}\right) + (1-\lambda) P_2(y,q)\log\left(\frac{ P_2(y,q)}{P_2(q)}\right)\\
&\ge \left[\lambda P_1(y,q)+ (1-\lambda) P_2(y,q)\right] \log \left(\frac{\lambda P_1(y,q) +  (1-\lambda) P_2(y,q)}{ \lambda P_1(q) + (1-\lambda) P_2(q)}\right).
\end{align*}
By summing over $(q,y)$ on both sides of the equation, the convexity of $-H(Y|Q)$ is concluded.

\section{BFC of type I --- Proof of Theorem \ref{theorem:Type1}}\label{app:type1}
In this section, we present the graph-based encoders and derive their achievable rates. First, the BCJR condition in \eqref{eq:BCJR}, for the BFC of type I, simplifies to:
\begin{align}
\label{eq:FadingI_BCJR}
 P_{S^+|Q^+}(0|g(q,y))
&=\left\{\begin{array}{cc}
 \frac{(1-p)\pi_{S|Q}(0|q)P_{X|S,Q}(0|0,q) + p\pi_{S|Q}(1|q)P_{X|S,Q}(1|1,q)}{\bar{p}+(2p-1)\pi_{S|Q}(1|q)P_{X|S,Q}(1|1,q)} & \text{if } y=0, \\
 \frac{p\left(1-\pi_{S|Q}(0|q)P_{X|S,Q}(0|0,q)-\pi_{S|Q}(1|q)P_{X|S,Q}(1|1,q)\right)}{p - (2p-1)\pi_{S|Q}(1|q)P_{X|S,Q}(1|1,q)} & \text{if } y=1, \end{array}\right.
\end{align}
where the property is satisfied if $P_{S^+|Q^+}(s|q) = \pi_{S|Q}(s|q)$ for all $(q,y)$.

\subsection{Proof of Theorem \ref{theorem:Type1}.1}
The graph-based encoder is based on a single-node $Q$-graph and the input distribution:
\begin{align}\label{eq:OR_EASY_distriubtion}
  P_{X|S,Q}(0|0,1) &= 0, \quad P_{X|S,Q}(1|1,1) = \frac{0.5}{1-p},
\end{align}
for $p\le0.5$. The induced stationary distribution is $\pi_{S|Q}(0|1) = p$.


Using \eqref{eq:FadingI_BCJR}, we now show that the BCJR-invariant property is satisfied for $(q=1,y=0)$:
\begin{align*}
  P_{S^+|Q^+}(0|1)
  &= \frac{p\bar{p}\frac{0.5}{\bar{p}}}{\bar{p}+(2p-1)\bar{p}\frac{0.5}{\bar{p}}} \\
  &= \pi_{S|Q}(0|1),
\end{align*}
and for $(q=1,y=1)$:
\begin{align*}
P_{S^+|Q^+}(0|1)
    &= \frac{p(1-\bar{p}\frac{0.5}{\bar{p}})}{p - (2p-1)\bar{p}\frac{0.5}{\bar{p}}} \\
    &= \pi_{S|Q}(0|1).
\end{align*}

Therefore, for $p\leq 0.5$, it follows that the achievable rate of the graph-based encoder is:
\begin{align*}
	R_1^\mathrm{I}(p) &= I(X,S;Y|Q) \\
           &= H_2\left(\bar{p}+(2p-1)\pi_{S|Q}(1|1)P_{X|S,Q}(1|1,1)\right)-H_2(p)\\
           &= 1-H_2(p).
\end{align*}

\subsection{Proof of Theorem \ref{theorem:Type1}.3}
The graph-based encoder is given by the $Q$-graph in Fig. \ref{fig:FC_I_Q2}, and the input distribution:
\begin{align}\label{eq:OR_EASY_distriubtion2}
    	P_{X|S,Q}(0|0,1) &= \frac{\bar{a}(2p - 1)}{p} \nn\\
        P_{X|S,Q}(1|1,1) &= 1 \nn\\
    	P_{X|S,Q}(0|0,2) &= \frac{p(p-a)(p\bar{a}+a\bar{p})}{\bar{a}(2p - 1)(p^2+a\bar{p}(2p-1))} \nn\\
    	P_{X|S,Q}(1|1,2) &= \frac{p^2(2a - 1)}{p-a(2p - 1)(p-\bar{a}(2p - 1))},
\end{align}
for all $p\in[0.5,1]$ and $a\in[0.5,1)$. In order to verify that  $0\preceq P_{X|S,Q}\preceq 1$, we need $(2p-1)^2a^2-\left(p+(2p-1)^2\right)a+p^2\ge 0$. Thus, there exists a graph-based encoder for any such $a$. The constraint does not appear in Theorem \ref{theorem:Type1}.3 since the maximum is unchanged due to this constraint. The computation of the achievable rate and the BCJR conditions are omitted here and follow from straightforward calculations.

\section{BFC of type II --- Proof of Theorem \ref{theorem:Type2_1}}\label{app:Type2_1}
\subsection{Proof of Theorem \ref{theorem:Type2_1}.1}
The proof is based on Theorem \ref{theorem:lower} with the $Q$-graph in Table \ref{table:FC_Q3}, and the input distribution:
\begin{align}\label{eq:OR_HARD_distriubtion}
  P_{X|S,Q}(0|1,1)&=0.5 \nn\\
  P_{X|S,Q}(0|1,2)&=P_{X|S,Q}(0|1,3)=0.
\end{align}

\begin{table}[h]
\caption{Q-graph for the BFC of type II (Theorem \ref{theorem:Type2_1}.1)}
\vspace{-1cm}
\label{table:FC_Q3}
\begin{center}
\begin{tabular}{ |c|c|c|c| } 	
 \hline
 $ $& $q^+=1$ & $q^+=2$ & $q^+=3$ \\
 \hline
 $q=1$ & $y=0$ & $y=1$ & $-$\\
  \hline
 $q=2$ & $-$ & $-$ & $y=0/1$\\
  \hline
 $q=3$ & $y=0$ & $y=1$ & $-$\\
 \hline
\end{tabular}
\bigskip\\
\end{center}
\vspace{-1cm}
\end{table}

It can now be verified that the BCJR property in \eqref{eq:BCJR} is satisfied. The verification and the computation of the achievable rate are omitted here, but follow by straightforward calculations.

\subsection{Proof of Theorem \ref{theorem:Type2_1}.2}
The proof is based on a verification of the KKT conditions for the suspected solution:
\begin{align}\label{eq:KKT_type2_solution}
P_{S,Q,X,Y}(0,1,0,0) &=  2P_{S,Q,X,Y}(1,3,1,1) = \frac{2p\bar{p}^2 }{D} \nn\\
P_{S,Q,X,Y}(0,1,0,1) &=  \left(\frac{2\bar{p}}{p}\right)P_{S,Q,X,Y}(0,3,1,1) = \frac{2p^2\bar{p}}{D}\nn\\
P_{S,Q,X,Y}(1,2,0,0) &=  P_{S,Q,X,Y}(1,2,1,1) = \frac{0.5p\bar{p}}{D}\nn\\
P_{S,Q,X,Y}(1,2,1,0) &=  P_{S,Q,X,Y}(1,2,0,1) = \frac{0.5p^2}{D}\nn\\
P_{S,Q,X,Y}(0,3,1,0) &=  P_{S,Q,X,Y}(1,3,1,0) = \frac{p^2\bar{p}}{D},
\end{align}
where $D \triangleq 2p(2 - p)$, and all the remaining entries of $P_{S,Q,X,Y}$ are equal to zero. The objective induced by the joint distribution in \eqref{eq:KKT_type2_solution} is equal to $R_1^\mathrm{II}(p)$ in Theorem \ref{theorem:Type2_1}.

Recall that for an optimization problem with an objective function $f(\cdot)$, equality constraints $h(\cdot)$ and inequality constraints $g(\cdot)$, the KKT conditions are:
\begin{itemize}
  \item Stationarity: $\bigtriangledown f(x^*) - \sum_i\mu_i\bigtriangledown g(x^*) - \sum_j\lambda_j\bigtriangledown h(x^*)=0$. \label{KKT:1}
  \item Feasibility: $g(x^*) \le 0$ and $h(x^*)=0$.\label{KKT:2}
  \item Dual feasibility: $\mu_i\ge0$ for all $i$.\label{KKT:3}
  \item Complementary slackness: $\mu_ig_i(x^*) = 0$ for all $i$.\label{KKT:4}
\end{itemize}
The following lemma shows that \eqref{eq:KKT_type2_solution} is indeed an optimal solution, for $p\ge p^*$:
\begin{lemma}[KKT conditions for \eqref{eq:KKT_type2_solution}] \label{lemma:KKT}
The following statements hold:
\begin{enumerate}
\item The feasibility constraints are satisfied for the solution in \eqref{eq:KKT_type2_solution}.
\item There exists a set of constants $\{\mu^*_i\}$ and $\{\lambda^*_j\}$ that solve the stationary constraints such that the complementary slackness conditions hold.
\item The inequality multipliers $\mu$ are non-negative iff $p\ge p^*$, where $p^*$ is the unique solution for $(2 + p)\log_2\left(\frac{2p^2}{1 - 2 p(1 - p)}\right) + \log_2\left(\frac{2(1 - p)^2}{1 - 2 p(1 - p)}\right)=0$.
\end{enumerate}
\end{lemma}

\noindent\emph{Proof of Lemma \ref{lemma:KKT}.1:} Straightforward calculation.\\
\emph{Proof of Lemma \ref{lemma:KKT}.2:} We will use three sets of KKT multipliers for the constraints:
\begin{enumerate}
  \item $\mu(s,q,x,y)$ for the inequality constraint $P(s,q,x,y)\ge0$.
  \item $\lambda_C(s,q,x,y)$ for the equality constraints of the channel.
  \item $\lambda_S(s,q)$ for the equality constraints of the stationary distribution.
  \item $\lambda_P$ for the equality constraint for the PMF.
\end{enumerate}
The stationarity constraints are satisfied with all $\mu(\cdot)$ variables that are equal to zero except:
\begin{align}
 \mu(1,1,0,0) &= \frac{(2p-1)[-(1+2p)(1 - \log(2p^2 - 2p + 1))  - 2\log(1 - p)  - 4p\log(p)]}{(1 - p)(2- p)}\label{eq:KKT_ineq_multi_1}\\
 \mu(1,1,1,0) &= \frac{(2p - 1 ) \left[(p + 3)(1-\log_2(1 - 2 p\bar{p})) + 2\log_2(1 - p) + 2 (2 + p) \log_2(p)\right]}{(-2 + p) p}.\label{eq:KKT_ineq_multi_2}
\end{align}
The other KKT multipliers are not presented and can be found explicitly by solving a set of linear equations that follow from the KKT stationary conditions.

\noindent\emph{Proof of Lemma \ref{lemma:KKT}.3}
First, we show the positivity of the first multiplier:
\begin{align}
 \mu(1,1,0,0) &= \frac{(2p-1)[-(1+2p)(1 - \log(2p^2 - 2p + 1))  - 2\log(1 - p)  - 4p\log(p)]}{(1 - p)(2- p)}\nn\\
 &\stackrel{(a)}\ge \frac{(2p-1)}{(1 - p)(2- p)}\left[\log_2\left(\frac{2p^2 - 2p + 1}{p^2 - 2p + 1}\right) - 1 + \frac{ 2p(1 - 2p)}{2p^2 - 2p + 1}\right]\nn\\
 &\stackrel{(b)}\ge 0,
\end{align}
where: $(a)$ follows from $- 2p\log\left(\frac{2p^2}{2p^2-2p+1}\right)\ge 2p\frac{1-2p}{2p^2-2p+1}$ and $(b)$ follows from the fact that the derivative (of the squared brackets) is an increasing function of $p$. 

For the other multiplier, $\mu(1,1,1,0)$, the denominator is always negative. Therefore, we concentrate on the squared brackets in \eqref{eq:KKT_ineq_multi_2}. The derivative of $(p + 3)(1-\log_2(1 - 2 p\bar{p}))$ and $2\log_2(1 - p)$ is negative for $p\ge0.6$. On the other hand, the function $2 (2 + p) \log_2(p)$ is increasing. Combining these two properties yields that the squared brackets cross the zero at one point above $p>0.5$.
By equating the squared brackets to zero and simplifying the expression, we get:
$(2 + p)\log_2\left(\frac{2p^2}{1 - 2 p(1 - p)}\right) +
 \log_2\left(\frac{2(1 - p)^2}{1 - 2 p(1 - p)}\right)=0$.
The point at which this equation equals zero is approximately $0.751$.

\section{BFC of type II --- Graph-based Encoders for $p\le0.5$}\label{app:Type2_2}
We define for some $a\in[0,1]$ the following quantities for the first graph-based encoder in Theorem \ref{theorem:type2_2}.1:
\begin{align}\label{eq:def_hard_LB_2}
  \Gamma_1(a,p) &= \bar{p}\nn\\
  \Gamma_2(a,p) &= 0.5\nn \\
  \Gamma_3(a,p) &= \bar{p}(2pa +\bar{a})\nn\\
  \Gamma_4(a,p) &= \frac{2p\bar{p}}{2pa +\bar{a}}\nn\\
  \Gamma_5(a,p) &= \frac{2p\bar{p}(1-2a) + a}{2(p + a - 3pa + 2p^2a)}\nn\\
  \Gamma_6(a,p) &= \frac{2p\bar{p}(2p^2 - 2p + 1)}{2p\bar{p}(1-2a) + a},
\end{align}
and the following quantities for the second graph-based encoder in Theorem \ref{theorem:type2_2}.2:
\begin{align}\label{eq:def_hard_LB_3}
  \Lambda_1(p) &= \bar{p}\nn\\
  \Lambda_2(p) &= 0.5\nn \\
  \Lambda_3(p) &= \frac{1 - 2p^2}{2\bar{p}}\nn\\
  \Lambda_4(p) &= \frac{1 + 2p\bar{p}(2p^2-6p+1)}{2\bar{p}(1-2p^2)}\nn\\
  \Lambda_5(p) &= \frac{- 8p^6 + 48p^5 - 116p^4 + 124p^3 - 58p^2 + 8p + 1}{8p^5 - 40p^4 + 60p^3 - 32p^2 + 2p + 2}\nn\\
  \Lambda_6(p) &= \frac{16p(1 - p)^7}{- 8p^6 + 48p^5 - 116p^4 + 124p^3 - 58p^2 + 8p + 1}.
\end{align}
\begin{theorem}\label{theorem:type2_2}
For the BFC of type II where $p\le0.5$:
\begin{enumerate}
\item There exists a graph-based encoder with $|\cQ|=6$ that achieves,
\begin{align}\label{eq:hard_LB_2}
    R_2^\mathrm{II}(p) &= \max_{0\le a\le 1}\sum_{q=1}^6 \pi_Q(q)H_2\left(\Gamma_q(a,p)\right) - H_2(p),
\end{align}
where
\begin{align*}
	\pi_Q&\triangleq  \frac1{N}[ \Gamma_3\Gamma_4 + \bar{\Gamma}_3\Gamma_5\Gamma_6 ,\  \bar{\Gamma}_1(\Gamma_3 + \bar{\Gamma}_3\Gamma_5), \ \bar{\Gamma}_1, \ \bar{\Gamma}_1\Gamma_3, \ \bar{\Gamma}_1\bar{\Gamma}_3, \ \bar{\Gamma}_1\bar{\Gamma}_3\Gamma_5],
\end{align*}
and $N$ is a normalization factor such that $\sum_q\pi_Q(q)=1$.
\item There exists a graph-based encoder with $|\cQ|=6$ that achieves,
\begin{align}\label{eq:hard_LB_3}
    R_3^\mathrm{II}(p) &= \sum_{q=1}^6 \tilde{\pi}_Q(q)H_2\left(\Lambda_q(p)\right) - H_2(p),
\end{align}
where,
\begin{align*}
	\tilde{\pi}_Q&\triangleq  \frac1{\tilde{N}}[ \Lambda_3 \Lambda_4 \Lambda_5 \Lambda_6, \ \bar{\Lambda}_1\Lambda_3 \Lambda_4 \Lambda_5, \ \bar{\Lambda}_1, \ \bar{\Lambda}_1\Lambda_3, \ \bar{\Lambda}_1\Lambda_3 \Lambda_4, \ \bar{\Lambda}_1\Lambda_3 \Lambda_4 \Lambda_5],
\end{align*}
and $\tilde{N}$ is a normalization factor such that $\sum_q\tilde{\pi}_Q(q)=1$.
\end{enumerate}
Therefore, $C_{\mathrm{II}}(p)\ge \max (R_2^\mathrm{II}(p),R_3^\mathrm{II}(p))$ for all $p\le0.5$. Note that the achievable rate in \eqref{eq:hard_LB_3} does not contain a maximization.
\end{theorem}

\archive{\subsection{Proof of Theorem \ref{theorem:type2_2}.1}\label{app:OR_hard_LB2}
\begin{table}[t]
\caption{Q-graph for the BFC type II (Theorem \ref{theorem:type2_2}.1).}
\vspace{-1cm}
\label{table:Tr_LB1_Q6}
\begin{center}
\begin{tabular}{ |c|c|c|c|c|c|c| } 	
 \hline
 $ $& $q^+=1$ & $q^+=2$ & $q^+=3$ & $q^+=4$ & $q^+=5$ & $q^+=6$ \\
 \hline
 $q=1$ & $y=0$ & $y=1$ & $-$ & $-$ & $-$ & $-$\\
 \hline
 $q=2$ & $-$ & $-$ & $y=0/1$ & $-$ & $-$ & $-$\\
 \hline
 $q=3$ & $-$ & $-$ & $-$ & $y=0$ & $y=1$ & $-$\\
 \hline
 $q=4$ & $y=0$ & $y=1$ & $-$ & $-$ & $-$ & $-$\\
 \hline
 $q=5$ & $-$ & $-$ & $y=1$ & $-$ & $-$ & $y=0$\\
 \hline
 $q=6$ & $y=0$ & $y=1$ & $-$ & $-$ & $-$ & $-$\\
 \hline
\end{tabular}
\end{center}
\end{table}
The proof is based on Theorem \ref{theorem:lower} with the $Q$-graph which is described in Table \ref{table:Tr_LB1_Q6}. The input distribution is given by:
\begin{align}\label{eq:Trapdoor_inputs2}
    P_{X|S,Q}(1|1,1) &= b\nn\\
	P_{X|S,Q}(1|1,2) &= \frac{1}{2}\nn\\
	P_{X|S,Q}(1|1,3) &= a\nn\\
    P_{X|S,Q}(1|1,4) &= 1\nn\\
    P_{X|S,Q}(1|1,5) &= \frac{a(1 - 2p\bar{p})}{2(a\bar{p} - ap\bar{p} + p^2)}\nn\\
    P_{X|S,Q}(1|1,6) &= 1,
\end{align}
for all $p\le0.5$, where $(a,b)\in[0,1]^2$. Straightforward calculation of the stationary distribution gives that:
\begin{align}
    	\pi_{S|Q}(0|1) &= 1 \nn\\
        \pi_{S|Q}(0|2) &= 0 \nn\\
        \pi_{S|Q}(0|3) &= p \nn\\
        \pi_{S|Q}(0|4) &= \frac{p(a + 1)}{\bar{a}+2pa} \nn\\
        \pi_{S|Q}(0|5) &= \frac{p\bar{p}\bar{a}}{p - a\bar{p}(2p - 1)} \nn\\
        \pi_{S|Q}(0|6) &= \frac{p(2pa - a - 4p + 2p^2 + 2)}{2p + a - 4pa + 4p^2a - 2p^2}.
\end{align}
It can now be verified that the BCJR-invariant property is satisfied and the expression $I(X,S;Y|Q)$ is the one, denoted by $R_2^\mathrm{II}(p)$, in Theorem \ref{theorem:type2_2}.
}
\subsection{Proof of Theorem \ref{theorem:type2_2}.2}\label{app:OR_hard_LB3}
\begin{table}[h!]
\caption{Q-graph for the BFC of type II (Theorem \ref{theorem:type2_2}.2)}
\vspace{-1cm}
\label{table:Tr_LB2_Q6}
\begin{center}
\begin{tabular}{ |c|c|c|c|c|c|c| } 	
 \hline
 $ $& $q^+=1$ & $q^+=2$ & $q^+=3$ & $q^+=4$ & $q^+=5$ & $q^+=6$ \\
 \hline
 $q=1$ & $y=0$ & $y=1$ & $-$ & $-$ & $-$ & $-$\\
 \hline
 $q=2$ & $-$ & $-$ & $y=0/1$ & $-$ & $-$ & $-$\\
 \hline
 $q=3$ & $-$ & $-$ & $y=1$ & $y=0$ & $-$ & $-$\\
 \hline
 $q=4$ & $-$ & $-$ & $y=1$ & $-$ & $y=0$ & $-$\\
 \hline
 $q=5$ & $-$ & $-$ & $y=1$ & $-$ & $-$ & $y=0$\\
 \hline
 $q=6$ & $y=0$ & $y=1$ & $-$ & $-$ & $-$ & $-$\\
 \hline
\end{tabular}
\end{center}
\end{table}
The proof is based on Theorem \ref{theorem:lower} with the $Q$-graph which is described in Table \ref{table:Tr_LB2_Q6}. The input distribution is given by:
\begin{align}\label{eq:Trapdoor_inputs3}
    P_{X|S,Q}(1|1,1) &= a\nn\\
	P_{X|S,Q}(1|1,2) &= \frac{1}{2}\nn\\
	P_{X|S,Q}(1|1,3) &= \frac{1 - 2p}{2\bar{p}^2}\nn\\
    P_{X|S,Q}(1|1,4) &= \frac{(2p - 1)^2}{2\bar{p}^2(1 - 2p\bar{p})}\nn\\
    P_{X|S,Q}(1|1,5) &= \frac{(1 - 2p)^3}{2\bar{p}^2(4p^4 - 12p^3 + 10p^2 - 4p + 1)}\nn\\
    P_{X|S,Q}(1|1,6) &= 1,
\end{align}
for all $p\le0.5$, where $a\in[0,1]$. Straightforward calculation of the stationary distribution gives that:
\begin{align}
    	\pi_{S|Q}(0|1) &= 1 \nn\\
        \pi_{S|Q}(0|2) &= 0 \nn\\
        \pi_{S|Q}(0|3) &= p \nn\\
        \pi_{S|Q}(0|4) &= \frac{p(2p^2 - 6p + 3)}{1 - 2p^2} \nn\\
        \pi_{S|Q}(0|5) &= \frac{p(4p^4 - 20p^3 + 38p^2 - 28p + 7)}{1 + 16p^3 - 14p^2 + 2p - 4p^4} \nn\\
        \pi_{S|Q}(0|6) &= \frac{p(8p^6 - 56p^5 + 164p^4 - 256p^3 + 214p^2 - 90p + 15)}{1 + 48p^5 - 116p^4 + 124p^3 - 58p^2 + 8p - 8p^6}.
\end{align}
It can now be verified that the BCJR-invariant property is satisfied and the expression $I(X,S;Y|Q)$ is the one, denoted by $R_3^\mathrm{II}(p)$, in Theorem \ref{theorem:type2_2}.

\section{The Ising channel --- Proof of Theorem \ref{theorem:Ising}}\label{app:Ising}
\begin{proof}
The proof is based on Theorem \ref{theorem:lower} with the $Q$-graph which is described in Table \ref{table:LB_Q6}. The input distribution is given by:
    \begin{align}\label{eq:ising_inputs}
    	P_{X|S,Q}(0|0,1) &= \frac{p(ba^2p - ba^2 + bap^2 - abp + ab - p^2)+k}{p^2(a - p - ab + abp)+k} \nn\\
        P_{X|S,Q}(0|1,1) &= \frac{\bar{a}(a - p - ab - ac - ap + cp + p^2 + 2abp + acp - abp^2)}{p^2(ab + p - a - abp)} \nn\\
    	P_{X|S,Q}(0|0,2) &= \frac{p(ab^2p - ab^2 + abp^2 - abp + ab - p^2)+k}{p^2(b - p - ab + abp)+k} \nn\\
    	P_{X|S,Q}(0|1,2) &= \frac{\bar{b}(a - p - ab - ac - ap + cp + p^2 + 2abp + acp - abp^2)}{p^2(ab + p - b - abp)} \nn\\
    	P_{X|S,Q}(0|0,3) &= \frac{p-c}{p} \nn\\
    	P_{X|S,Q}(0|1,3) &= \frac{\bar{c}p(a - p - ab + abp)}{a - p - ab - ac - ap + cp + p^2 - p^3 + 3abp + acp - 3abp^2 + abp^3} \nn\\
    	P_{X|S,Q}(0|0,4) &= P_{X|S,Q}(1|1,3)\nn\\
        P_{X|S,Q}(0|1,4) &= P_{X|S,Q}(1|0,3) \nn\\
    	P_{X|S,Q}(0|0,5) &= P_{X|S,Q}(1|1,1)\nn\\
        P_{X|S,Q}(0|1,5) &= P_{X|S,Q}(1|0,1) \nn\\
    	P_{X|S,Q}(0|0,6) &= P_{X|S,Q}(1|1,2)\nn\\
        P_{X|S,Q}(0|1,6) &= P_{X|S,Q}(1|0,2),
    \end{align}
for all $p\ge0.5$, where $(a,b,c)\in[0,1]^3$ are parameters that are restricted to $\cV \triangleq \{(a,b,c): 0\preceq P_{X|S,Q}\preceq 1\}$ and $k \triangleq 1-\bar{p}[(1+2p^2)(1-ab)+a\bar{c}\bar{p}+cp+3abp]$.

Straightforward calculation of the stationary distribution gives that:
\begin{align}
    	\pi_{S|Q}(0|1) &= a \nn\\
        \pi_{S|Q}(0|2) &=\frac{b - \bar{p}(a - \bar{c})}{a(2p - 1) + \bar{p}(b +\bar{c})}\nn \nn\\
        \pi_{S|Q}(0|3) &= \frac{p(a -\bar{c})}{\bar{p}(b + \bar{c}) + a(2p - 1) - 1}\nn \\
        \pi_{S|Q}(0|4) &= \pi_{S|Q}(1|3) \nn\\
        \pi_{S|Q}(0|5) &= \pi_{S|Q}(1|1)\nn \\
        \pi_{S|Q}(0|6) &= \pi_{S|Q}(1|2).\nn
\end{align}

The BCJR equation in \eqref{eq:BCJR} can be written explicitly for the Ising channel as follows:
\begin{align}
\label{eq:Ising_BCJR}
 P_{S^+|Q^+}(0|g(q,y))
&=\left\{\begin{array}{cc}
 \frac{\delta_q+\bar{p}(\pi_{S|Q}(1|q)-\gamma_q)}{\bar{p}(1+\delta_q-\gamma_q)+(2p-1)\pi_{S|Q}(0|q)} & \text{if } y=0, \\
 \frac{p(\pi_{S|Q}(1|q)-\gamma_q)}{1-\bar{p}(1+\delta_q-\gamma_q)-(2p-1)\pi_{S|Q}(0|q)} & \text{if } y=1, \end{array}\right.
\end{align}
for all $(q,y)$, where $\delta_q \triangleq \pi_{S|Q}(0|q)P_{X|S,Q}(0|0,q)$ and $\gamma_q\triangleq \pi_{S|Q}(1|q)P_{X|S,Q}(1|1,q)$.

\begin{table}[t]
\caption{Q-graph for the Ising channel (Theorem \ref{theorem:Ising})}
\vspace{-1cm}
\label{table:LB_Q6}
\begin{center}
\begin{tabular}{ |c|c|c|c|c|c|c| } 	
 \hline
 $ $& $q^+=1$ & $q^+=2$ & $q^+=3$ & $q^+=4$ & $q^+=5$ & $q^+=6$ \\
 \hline
 $q=1$ & $-$ & $y=0$ & $y=1$ & $-$ & $-$ & $-$\\
 \hline
 $q=2$ & $y=0$ & $-$ & $y=1$ & $-$ & $-$ & $-$\\
 \hline
 $q=3$ & $-$ & $-$ & $-$ & $y=0$ & $y=1$ & $-$\\
 \hline
 $q=4$ & $y=0$ & $-$ & $y=1$ & $-$ & $-$ & $-$\\
 \hline
 $q=5$ & $-$ & $-$ & $-$ & $y=0$ & $-$ & $y=1$\\
 \hline
 $q=6$ & $-$ & $-$ & $-$ & $y=0$ & $y=1$ & $-$\\
 \hline
\end{tabular}
\end{center}
\end{table}
It can now be verified that the BCJR-invariant property is satisfied and the expression $I(X,S;Y|Q)$ is the one in Theorem \ref{theorem:Ising}. Finally, instead of computing the region $\cV$ explicitly, we defined a subset set $\cL\subseteq\cV$ in \eqref{eq:ising_L} such that the lower bound performance is unchanged. The inclusion of $\cL\subset\cV$ is omitted here and follows from tedious technical arguments.
\end{proof}

\section{The trapdoor channel --- Proof of Theorem \ref{theorem:TRAPDOOR}}\label{app:trapdoor}
\subsection{Proof of Theorem \ref{theorem:TRAPDOOR}.1}\label{app:trapdoor_LB}
\begin{figure}[t]
\centering
    \psfrag{E}[][][1]{$Y=0$}
    \psfrag{D}[][][1]{$Y=1$}
    \psfrag{C}[][][1]{$Q=3$}
    \psfrag{B}[][][1]{$Q=2$}
    \psfrag{A}[][][1]{$Q=1$}
    \includegraphics[scale = 0.5]{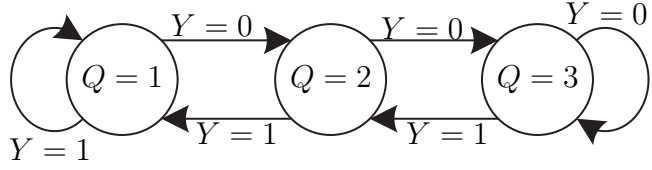}
    \caption{$Q$-graph for the Trapdoor channel.}
    \label{fig:Q1_trapdoor}
\end{figure}

The proof is based on Theorem \ref{theorem:lower} and the $Q$-graph from Fig. \ref{fig:Q1_trapdoor}. The input distribution is given by:
\begin{align*}
    P_{X|S,Q}(0|0,1) &= 1 \\
    P_{X|S,Q}(1|1,1) &= \frac{p}{\bar{p}} \\
    P_{X|S,Q}(0|0,2) &= \frac1{2\bar{p}} \\
    P_{X|S,Q}(1|1,2) &= \frac1{2\bar{p}} \\
    P_{X|S,Q}(0|0,3) &= \frac{p}{\bar{p}} \\
    P_{X|S,Q}(1|1,3) &= 1.
\end{align*}
where $p\le0.5$ is the channel parameter. Straightforward calculation of the stationary distribution gives that $[\pi_{S|Q}(0|1),\pi_{S|Q}(0|2),\pi_{S|Q}(0|3)] = [\frac{1-2p}{2(1-p)}, 0.5, \frac{1}{2(1-p)}]$.

The BCJR equation in \eqref{eq:BCJR} can be written explicitly for the trapdoor channel as follows:
\begin{align}
\label{eq:TRAPDOOR_BCJR}
 P_{S^+|Q^+}(0|g(q,y))
&=\left\{\begin{array}{cc}
 \frac{\delta_q}{(1-p)(\delta_q-\gamma_q + \pi_{S|Q}(1|q)) + p\pi_{S|Q}(0|q)} & \text{if } y=0, \\
 \frac{\bar{p}(\pi_{S|Q}(0|q)-\delta_q) + p(\pi_{S|Q}(1|q) -\gamma_q)}{1 - \bar{p}(\delta_q-\gamma_q + \pi_{S|Q}(1|q)) - p\pi_{S|Q}(0|q)} & \text{if } y=1, \end{array}\right.
\end{align}
for all $(q,y)$, where $\delta_q \triangleq \pi_{S|Q}(0|q)P_{X|S,Q}(0|0,q)$ and $\gamma_q\triangleq \pi_{S|Q}(1|q)P_{X|S,Q}(1|1,q)$.

The BCJR-invariant property is verified for $q=1$ as follows:
\begin{align*}
     P_{S^+|Q^+}(0|1)&=\frac{\delta_1}{(1-p)(\delta_1-\gamma_1 + \pi_{S|Q}(1|1)) + p\pi_{S|Q}(0|1) }\\
     &= \frac{\frac{1-2p}{2(1-p)}}{(1-p)(1 - \frac{p}{2(1-p)^2})
 + p\frac{1-2p}{2(1-p)}} \\
        &= 0.5\\
        &=\pi_{S|Q}(0|1);\\
 P_{S^+|Q^+}(0|2)&=\frac{(1-p)(\pi_{S|Q}(0|2)-\delta_2) + p(\pi_{S|Q}(1|2) -\gamma_2)}{1 - (1-p)(\delta_2-\gamma_2 + \pi_{S|Q}(1|2)) - p\pi_{S|Q}(0|2)}\\
 &= \frac{p(\frac1{2(1-p)} -\frac{p}{2(1-p)^2})}{1 - (1-p)(1 - \frac{p}{2(1-p)^2})
 - p\frac{1-2p}{2(1-p)}}\\
 &= \frac{1-2p}{2(1-p)}\\
 &= \pi_{S|Q}(0|2).
\end{align*}
The verification of \eqref{eq:TRAPDOOR_BCJR} for the other edges can be done in a similar way.

Calculation of the transition probabilities gives that:
\begin{align*}
     P_{Q^+|Q}(1|1)&= P_{Q^+|Q}(1|1) = \frac{1-2p}{1-p}\\
     P_{Q^+|Q}(1|2)&= P_{Q^+|Q}(3|2) = 0.5,
\end{align*}
which gives that the stationary distribution is
\begin{align*}
  [\pi_Q(1),\pi_Q(2),\pi_Q(3)]&=\left[\frac{1-p}{2(2-3p)},\frac{1-2p}{2-3p},\frac{1-p}{2(2-3p)}\right],
\end{align*}
and the per node rewards are
\begin{align*}
  I(X,S;Y|Q=i)&= H_2\left(\frac{1-2p}{1-p}\right) - H_2(p)\frac{1-2p}{2(1-p)^2} \\
  I(X,S;Y|Q=2)&= 1 - H_2(p) \frac{1-2p}{2(1-p)},
\end{align*}
for $i=1,3$.

The corresponding achievable rate can then be computed as follows:
\begin{align*}
  C_\mathrm{T}(p)&\geq  I(X,S;Y|Q) \nn\\
               &= 2 \frac{1-p}{2(2-3p)} \left(H_2\left(\frac{1-2p}{1-p}\right) - H_2(p)\frac{1-2p}{2(1-p)^2}\right) + \frac{1-2p}{2-3p}\left(1 - H_2(p) \frac{1-2p}{2(1-p)}\right)\nn\\
               &= \frac{1-p}{2-3p} H_2\left(\frac{1-2p}{1-p}\right) + \frac{1-2p}{2-3p} - H_2(p)\frac{1-2p}{2-3p}\\
               &= \frac{H_2(z) + z\left(1-H_2\left(\frac{1-z}{2-z}\right)\right)}{1+z} ,
\end{align*}
where $z(p)\triangleq \frac{1-2p}{1-p}$.

\subsection{Proof of Theorem \ref{theorem:TRAPDOOR}.2}\label{app:trapdoor_LB2}
\begin{table}[t]
\caption{Q-graph for the trapdoor channel (Theorem \ref{theorem:TRAPDOOR}.2)}
\vspace{-1cm}
\label{table:Trapdoor_Q4}
\begin{center}
\begin{tabular}{ |c|c|c|c|c| } 	
 \hline
 $ $& $q^+=1$ & $q^+=2$ & $q^+=3$ & $q^+=4$ \\
 \hline
 $q=1$ & $y=0$ & $y=1$ & $-$ & $-$\\
 \hline
 $q=2$ & $-$ & $-$ & $y=0$ & $y=1$\\
 \hline
 $q=3$ & $y=0$ & $y=1$ & $-$ & $-$\\
 \hline
 $q=4$ & $-$ & $-$ & $y=0$ & $y=1$\\
 \hline
\end{tabular}
\end{center}
\end{table}
The proof is based on Theorem \ref{theorem:lower} with the $Q$-graph described in Table \ref{table:Trapdoor_Q4}. The input distribution is given by:
\begin{align}\label{eq:Trapdoor_inputs}
    P_{X|S,Q}(0|0,1) &= \frac{b}{a} \nn\\
    P_{X|S,Q}(1|1,1) &=\frac{a\bar{a}\bar{p}-\bar{a}b+a^2p-abp}{a\bar{a}\bar{p}} \nn\\
    P_{X|S,Q}(0|0,2) &= \frac{(a\bar{a}\bar{p}-\bar{a}b+a^2p-abp)(a\bar{p}^2+2ab\bar{p}^2+5a^2p\bar{p}-b\bar{p}-a^2\bar{p}-abp\bar{p}+bp^2-a^2)}{p\bar{p}a\bar{a}(bp - ab\bar{p} - 2a^2p + a^2)} \nn\\
    P_{X|S,Q}(1|1,2) &= \frac{(b - a)(a\bar{p}^2+2ab\bar{p}^2+5a^2p\bar{p}-b\bar{p}-a^2\bar{p}-abp\bar{p}+bp^2-a^2-a\bar{a}p\bar{p})}{p\bar{a}(a\bar{a}\bar{p}-\bar{a}b+a^2p-abp)} \nn\\
    P_{X|S,Q}(0|0,3) &= P_{X|S,Q}(1|1,2) \nn\\
    P_{X|S,Q}(1|1,3) &= P_{X|S,Q}(0|0,2) \nn\\
    P_{X|S,Q}(0|0,4) &= P_{X|S,Q}(1|1,1) \nn\\
    P_{X|S,Q}(1|1,4) &= P_{X|S,Q}(0|0,1),
\end{align}
for all $p>0.5$, where $(a,b)$ are parameters that are restricted to $\cV \triangleq \{(a,b): 0\preceq P_{X|S,Q}\preceq 1\}$. In Theorem \ref{theorem:TRAPDOOR}.2 we mention a single constraint on the parameters $(a,b)$, since we present the optimal encoder and the maximum is unchanged from the other constraints that appear due to the restriction $\cV \triangleq \{(a,b): 0\preceq P_{X|S,Q}\preceq 1\}$

Straightforward calculation of the stationary distribution gives that:
\begin{align}
    	\pi_{S|Q}(0|1) &= a \nn\\
        \pi_{S|Q}(0|2) &= \frac{pb - ab - 2pa^2 + a^2 + pab}{(a - b)\bar{p}} \nn\\
        \pi_{S|Q}(0|3) &= \pi_{S|Q}(1|2) \nn\\
        \pi_{S|Q}(0|4) &= \pi_{S|Q}(1|1).
\end{align}
The BCJR-invariant property is given by equation \eqref{eq:TRAPDOOR_BCJR}.
It can now be verified that the BCJR-invariant property is satisfied and the expression $I(X,S;Y|Q)$ is the one, denoted by $R_\mathrm{T_2}(p)$, in Theorem \ref{theorem:TRAPDOOR}.
\subsection{Proof of Theorem \ref{theorem:TRAPDOOR}.3 (Upper bound)}\label{app:trapdoor_UB}
From Theorem \ref{theorem:UB} with the $Q$-graph from Fig. \ref{fig:Q1_trapdoor}, we have
\begin{align*}
  D &\triangleq 2(-(1 + (-1 + p)p_2) (-1 + p_3) + p_1 (1 - p_2 + p (1 + p_2 - p_3) + p_3))\\
  \pi_1&= \pi_6=\frac{\bar{p_3}(1 - \bar{p}p_2)}{D} \\
  \pi_2&= \pi_5=\frac{pp_1 + \bar{p}p_1p_3 }{D}\\
  \pi_3&= \pi_4=\frac{p_1(1- \bar{p}p_2)}{D},
\end{align*}
where $p_1,p_2,p_3$ are parameters in $[0,1]$. The resultant upper bound is then
\begin{align*}
  C_\mathrm{T}(p) &\le \max 2(\pi_1+\pi_2)\left[H_2\left(\frac{\pi_1(1-\bar{p}p_1) + \pi_2\bar{p}p_2}{\pi_1+\pi_2}\right) - H_2(p)\left(\frac{p_1\pi_1 + p_2\pi_2}{\pi_1+\pi_2}\right)\right] \\
  & \ \ + (\pi_3+\pi_4)\left[H_2\left(\frac{\pi_3(1-\bar{p}p_3) + \pi_4\bar{p}p_4}{\pi_3+\pi_4}\right) - H_2(p)\left(\frac{p_3\pi_3 + p_4\pi_4}{\pi_3+\pi_4}\right)\right]\\
    &\stackrel{(a)}= \max 2(\pi_1+\pi_2)H_2\left(\frac{\pi_1\bar{p}p_1 + \pi_2(1-\bar{p}p_2)}{\pi_1+\pi_2}\right) - 2 H_2(p)(p_1\pi_1 + p_2\pi_2+p_3\pi_3)  + 2\pi_3\\
    &\stackrel{(b)}= \max 2(\pi_1+\pi_2)H_2\left(\frac{0.5-(\pi_1+\pi_2)}{\pi_1+\pi_2}\right) - 2 H_2(p)(p_1\pi_1 + p_2\pi_2+p_3\pi_3) + 2\pi_3 \\
    &\stackrel{(c)}= \max 2(\pi_1+\pi_2)H_2\left(\frac{0.5-(\pi_1+\pi_2)}{\pi_1+\pi_2}\right) - 2 H_2(p)\left(\pi_3 + \frac{p_1p_2(p+\bar{p}p_3)}{D}\right) + 2\pi_3 \\
    &\le \max 2(\pi_1+\pi_2)H_2\left(\frac{0.5-(\pi_1+\pi_2)}{\pi_1+\pi_2}\right) - 2 H_2(p)\pi_3 + 2\pi_3 \\
    &\stackrel{(d)}= \max_a \frac{H_2\left(a\right)+a(1-H_2(p))}{1+a}.
\end{align*}
where $(a)$ follows from $H_2(x) = H_2(1-x)$ and $p_3=p_4$, $(b)$ follows from $\pi_1\bar{p}p_1 + \pi_2(1-\bar{p}p_2) = \pi_3$ and $\pi_1+\pi_2+\pi_3=0.5$, $(c)$ follows from explicit calculation of $(p_1\pi_1 + p_2\pi_2+p_3\pi_3)$, and, finally, $(d)$ is due to the notation $a\triangleq \frac{0.5-(\pi_1+\pi_2)}{\pi_1+\pi_2}$ which implies that $2\pi_3 = \frac{a}{1+a}$, and by taking a maximization with respect to $a$.
\subsection{Proof of Theorem \ref{theorem:TRAPDOOR}.4 (Capacity is achieved at a single point)}
We show that there exists $p^*$ such that the lower and the upper bounds coincide. Towards this, we show that $ z(p)\triangleq \frac{1-2p}{1-p}$ achieves the maximum in the upper bound at a unique point. First, by calculating the derivative of \eqref{eq:Trapdoor_UB}, the maximizing argument, $a^*$, is the unique solution of $\log\left(\frac{(1-a)^2}{a}\right) = 1-H_2(p)$. Then, by substituting $z(p)\triangleq \frac{1-2p}{1-p}$ into this equation we get $1 + 2 \log\frac{p}{1 - p} - \log\frac{1 - 2 p}{1 - p} -H_2(p) = 0$. It can be shown that the LHS is an increasing function when $p<0.5$, that crosses the zero at one point only.

\section*{Acknowledgment}
The authors would like to thank Dr. Or Ordentlich for suggesting the study of the binary fading channels in Section \ref{sec: BFCs}.
\bibliography{ref}
\bibliographystyle{IEEEtran}
\end{document}